\documentclass[preprint]{aastex63}
\usepackage{bbold}
\usepackage{bbm}
\usepackage{times}
\usepackage{amsmath,amssymb}
\submitjournal{ApJ}

\shorttitle{Escaping the lava worlds}
\shortauthors{Kang et al.}

\begin{document}

\title{Comet-like tails of disintegrating exoplanets explained by escaping outflows emanated from the permanent nightside: day-side versus night-side escape} 

\correspondingauthor{Wanying Kang}
\email{wanying@mit.edu}

\author[0000-0002-4615-3702]{Wanying Kang}
\affil{Earth, Atmospheric and Planetary Science Department, 
  Massachusetts Institute of Technology, 
  Cambridge, MA 02139, USA}

\author[0000-0001-7758-4110]{Feng Ding}
\affil{School of Engineering and Applied Sciences, 
  Harvard University, 
  Cambridge, MA 02138, USA}

\author[0000-0003-1127-8334]{Robin Wordsworth}
\affil{School of Engineering and Applied Sciences, 
  Harvard University, 
  Cambridge, MA 02138, USA}

\author[0000-0002-6892-6948]{Sara Seager}
\affil{Earth, Atmospheric and Planetary Science Department, 
  Massachusetts Institute of Technology, 
  Cambridge, MA 02139, USA}

\begin{abstract}
Ultra-hot disintegrating exoplanets have been detected with tails trailing behind and/or shooting ahead of them. These tails are believed to be made of dusts that are carried upward by the supersonic flow escaping the planet's gravity field from the fiercely heated permanent day-side. Conserving angular momentum, this day-side escape flux would lead the planet in orbit. In order to explain the trailing tails in observation, radiation pressure, a repulsive force pushing the escape flow away from the host star is considered to be necessary.

We here investigate whether escape could occur on the night-side as the escape flow is deflected by the pressure gradient force. We demonstrate in an idealized framework that escape flux from the night-side could dominate that from the day-side; and the former may naturally explain the commonly-observed trailing tails based on angular momentum conservation, without the need to invoke radiation pressure, which has previously been thought to be the key. We also find analytical approximations for both dayside and nightside escape fluxes, which may be applied to study planetary evolution of disintegrating planets and to infer planetary sizes from observations of the properties of their dusty tails. 
\end{abstract}


\section{Introduction}
Tidally-locked exoplanets, orbiting extremely close to their host stars, receive fierce radiation on the permanent day-side that can melt their rocky surface to form a lava ocean there. Owing to their short orbital periods, many lava planets have been detected \citep{Henning-Renaud-Saxena-et-al-2018:highly, Leger-Grasset-Fegley-et-al-2011:extreme, Hammond-Pierrehumbert-2017:linking, Rouan-Deeg-Demangeon-et-al-2011:orbital}, and these include relatively small planets \citep{Rappaport-Levine-Chiang-et-al-2012:possible, Brogi-Keller-Juan-et-al-2012:evidence, Rappaport-Barclay-DeVore-et-al-2014:koi, Sanchis-Ojeda-Rappaport-Palle-et-al-2015:the, Budaj-Kocifaj-Salmeron-et-al-2015:tables}, whose radial velocity signature is below the measurement noise level \citep{Lieshout-Rappaport-2018:disintegrating}.

Unexpectedly, the transit depth of these planets seem to vary from orbit to orbit and the transit curve is asymmetry between the ingress and egress. After excluding the possibility of binary star system, dual planet system, \citet{Rappaport-Levine-Chiang-et-al-2012:possible} proposes that the tail is made of dust carried by the escape flow from the planet. Because of their weak gravity, these small lava planets (e.g., KIC-1255b, KOI-2700b and K2-22b), cannot hold onto the mineral vapor evaporated from the lava ocean. The outgoing escape flow could bring aerosols from the surface, which would gradually melt under the stellar radiation, forming dusty tails that are weakening away from the planet \citep{Rappaport-Levine-Chiang-et-al-2012:possible, Lieshout-Min-Dominik-2014:dusty, Rappaport-Barclay-DeVore-et-al-2014:koi, Sanchis-Ojeda-Rappaport-Palle-et-al-2015:the}. This disintegrating process provides a unique chance to probe the chemical composition of the planet, and meanwhile the mineral vapor escape by itself is an interesting problem.

It is puzzling why KIC-1255b and KOI-2700b have tails lagged behind them in orbit \citep{Rappaport-Levine-Chiang-et-al-2012:possible, Brogi-Keller-Juan-et-al-2012:evidence, Rappaport-Barclay-DeVore-et-al-2014:koi}. Intuitively, one would expect vaporization and atmospheric escape to occur on the day-side of the planets, where the surface is directly heated by the stellar radiation. Then by conserving angular momentum, this day-side escape flow would overtake the planet in orbit and form a leading tail, rather than the trailing tails as observed \citep{Brogi-Keller-Juan-et-al-2012:evidence, Rappaport-Barclay-DeVore-et-al-2014:koi, Sanchis-Ojeda-Rappaport-Palle-et-al-2015:the, Budaj-Kocifaj-Salmeron-et-al-2015:tables}, unless there is a repulsive force pushing the escape flow away from the host star. \citet{Lieshout-Min-Dominik-2014:dusty, Rappaport-Barclay-DeVore-et-al-2014:koi} and \citet{Sanchis-Ojeda-Rappaport-Palle-et-al-2015:the} suggest that radiation pressure can serve such a purpose \footnote{Stellar wind also contributes, but is negligible compared to radiation pressure \citep{Rappaport-Barclay-DeVore-et-al-2014:koi}.}. While absorbing photons from the star, dust particles \cite[particularly those around 1$\mu$m][]{Lieshout-Min-Dominik-2014:dusty} inherit the photon's momentum, accelerating away from the star, and hence lagging behind the planet.
In this work, we will investigate whether the pressure gradient force alone can deflect the flow from day-side to night-side of the planet, and as a result, escape occurs on the night-side despite the low surface temperature there and trails behind the planet by virtue of angular momentum conservation. The other question is under what conditions (planet’s size and temperature), escape flux from the day-side would dominate over that from the night-side and vice versa. Using this, we may be able to constrain the remarkable uncertainty of the planetary size through observations of its tails. The focus of our work will be put on the escape process, i.e. how fast mineral vapor will vaporize from magma ocean and escape the planetary gravity field from day-side and night-side, under the constraint of condensation. This is complementary to the previous studies focusing on the trajectory and revaporization of dust particles after they have escaped the planetary gravity \citep{Lieshout-Min-Dominik-2014:dusty, Rappaport-Barclay-DeVore-et-al-2014:koi, Sanchis-Ojeda-Rappaport-Palle-et-al-2015:the}.

The key to the above questions is to distinguish the day-side and night-side escape, which has been overseen by traditional 1D escape models by assuming isotropy. In our framework, atmospheric pressure gradient between the day-side and night-side replaces the radiation pressure in \citet{Lieshout-Min-Dominik-2014:dusty, Rappaport-Barclay-DeVore-et-al-2014:koi} and \citet{Sanchis-Ojeda-Rappaport-Palle-et-al-2015:the}, diverting the flow toward the night-side (away from the star). The extremely low surface temperature on the night-side drives the pressure there close to zero, making the night-side as ``attractive'' a destination as the vacuum space. On the way toward the night-side, vapor flow would turn supersonic; and once that occurs, information of air pressure, temperature etc. cannot propagate upstream to exert impacts on the upstream flow over the day-side. As a result, mass would be continuously transported toward the night-side at a rate that is purely determined by the conditions on the day-side. Air mass from day-side accumulates near the anti-stellar point, making it a singular point of pressure, which in turn pumps mineral vapor upward from the night-side of the planet, away from the host star.

Flow field like this has been seen in multidimensional hydrodynamic simulations for hydrogen escape \citep{Tripathi-Kratter-Murray-Clay-et-al-2015:simulated, Shaikhislamov-Khodachenko-Lammer-et-al-2018:aeronomy, Debrecht-Carroll-Nellenback-Frank-et-al-2019:photoevaporative, McCann-Murray-Clay-Kratter-et-al-2019:morphology, Stone-Proga-2009:anisotropic}. Even without radiation pressure or stellar wind, a significant proportion of atmosphere is transported toward the night-side by the pressure gradient force, from where atmosphere escapes \citep{Tripathi-Kratter-Murray-Clay-et-al-2015:simulated, Shaikhislamov-Khodachenko-Lammer-et-al-2018:aeronomy, Debrecht-Carroll-Nellenback-Frank-et-al-2019:photoevaporative}. Although there are new processes (such as condensation and mass exchange between atmosphere and lava ocean) that do not exist on hot Jupiter to be considered for lava planets, we expect the general physical picture to remain qualitatively similar, except that the transport from day-side to night-side may be further enhanced by the strong pressure gradient on lava planets\footnote{The pressure on the night-side is not zero on hot Jupiters with a hydrogen envelope, but should be close to zero on lava planets in absence of the transport flow. This difference makes it difficult to compare our results with the previous multidimensional hydrodynamic simulations.}.

Simulations of the aforementioned escape flow are beyond the capabilities of a traditional 1D hydrodynamic escape framework, where isotropy is assumed \citep{Parker-1965:dynamical, Watson-Donahue-Walker-1981:dynamics, Perez-Becker-Chiang-2013:catastrophic, Lehmer-Catling-Zahnle-2017:longevity, Zahnle-Catling-2017:cosmic, Lammer-Kasting-Chassefiere-et-al-2008:atmospheric, Owen-2019:atmospheric}. Multidimensional hydrodynamic calculation has been utilized for simulating hydrogen escape on hot Jupiters \citep{Tripathi-Kratter-Murray-Clay-et-al-2015:simulated, Shaikhislamov-Khodachenko-Lammer-et-al-2018:aeronomy, Debrecht-Carroll-Nellenback-Frank-et-al-2019:photoevaporative, Wang-Dai-2018:evaporation, Khodachenko-Shaikhislamov-Lammer-et-al-2019:global}, but has not been applied to mineral vapor escape from small lava planets. Numerically, it could be challenging to deal with phase changes, mass exchange between mineral vapor atmosphere and different surface types (lava ocean and solidified surface), as well as orders of magnitude of pressure variation in both horizontal and vertical direction in a multidimensional hydrodynamic model. These motivate us to build an idealized theoretical framework that can distinguish the day-side and night-side escape. 

\section{Method brief}
\label{sec:method-brief}

 \begin{figure*}[hptb!]
  \centering
  \includegraphics[width=0.75\textwidth]{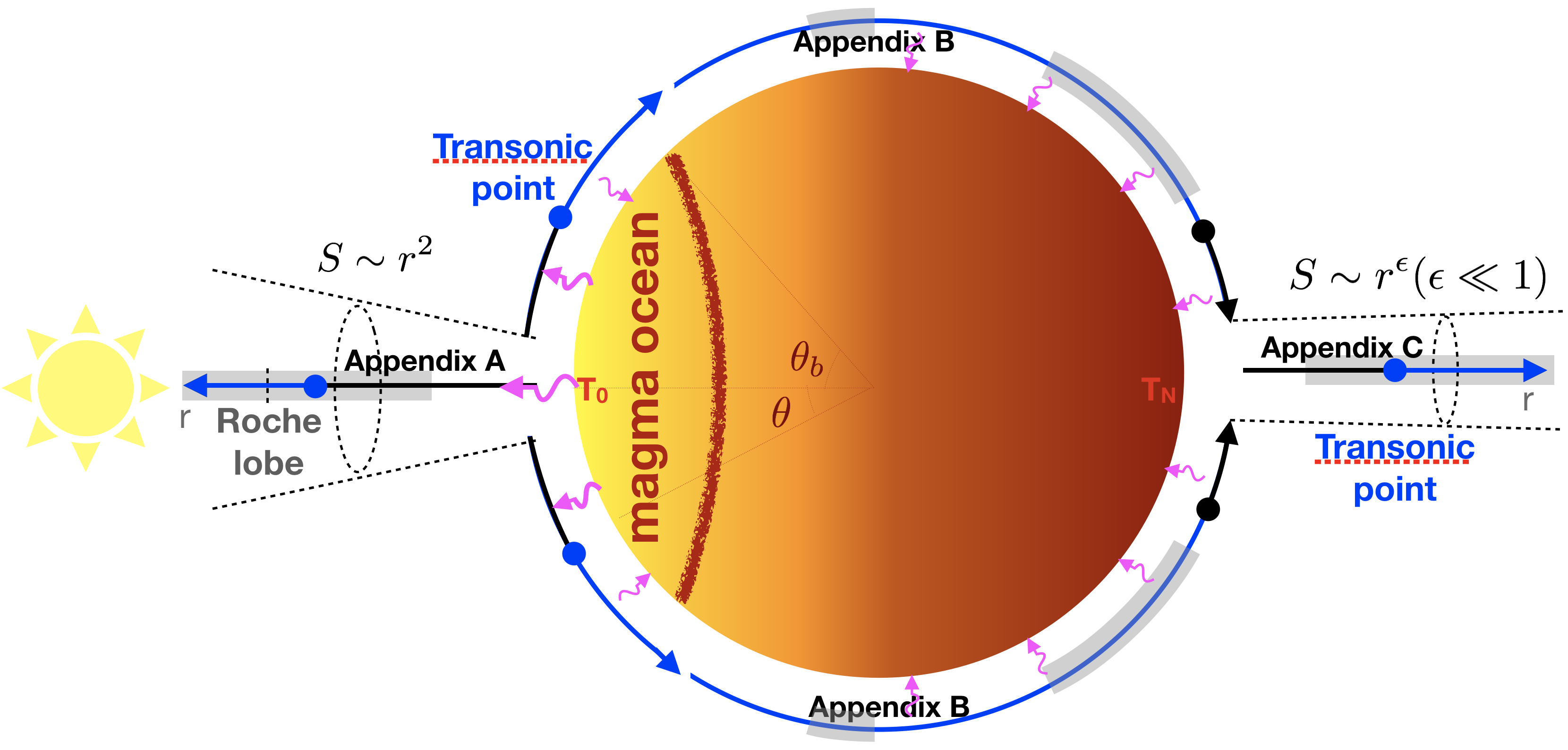}
  \caption{Model schematics. Both day-side escape and the two stages of night-side escape are considered here. We use arrows to represent the flow during each stage of escape, with corresponding section numbers marked on the side. Blue (black) arrows denote supersonic (subsonic) flow and blue (black) dots denote the locations where a subsonic (supersonic) flow turns into a supersonic (subsonic) flow. Gray shadings denote saturation and condensation. Magenta curly arrows denote surface exchange flux. Escape cross section is illustrated by a cone in dashed lines. $r$ is the distance from the center of the planet. $\theta$ is the tidally locked latitude counted from the substellar point, and $\theta_b$ denotes the latitude boundary of magma ocean. }
  \label{fig:schematics}
\end{figure*}

As sketched in Fig.~\ref{fig:schematics}, we combine three 1D models to depict the aforementioned anisotropic escape flow: 1) a day-side hydrodynamic escape model solving for the escape rate driven by the pressure gradient between the day-side surface and the vacuum space (appendix~\ref{sec:appendix-vertical-escape-day}), 2) a horizontal transport model calculating the day-side to night-side mass transport following the surface induced by the pressure gradient force (appendix~\ref{sec:appendix-horizontal-transport}), and 3) a night-side hydrodynamic escape model dealing with the escape process of the mass transported from the day-side (appendix~\ref{sec:appendix-vertical-escape-night}). 2) and 3) are connected by conserving energy and mass flux. Phase change is calculated explicitly in all three components, rather than being prescribed as in previous works, e.g., \citet{Perez-Becker-Chiang-2013:catastrophic}. Explicit consideration of phase change is necessary, because, unlike an atmosphere that is purely composed of hydrogen or water vapor, which remains unsaturated or saturated throughout, the mineral gas could undergo multiple transitions between subsaturated and saturated states depending on its geometry and external energy sources (regions that could be saturated are illustrated by gray shading in Fig.~\ref{fig:schematics}).

The two vertical escape models are built upon \citet{Lehmer-Catling-Zahnle-2017:longevity} and the horizontal transport model is built upon \citet{Ingersoll-Summers-Schlipf-1985:supersonic} to explicitly accounts for the transitions between saturation and undersaturation. Both escape and transport models solve temperature, pressure and velocity profiles from a set of equations given by mass continuity, momentum theorem and energy conservation, assuming steady state. The governing equations are singular around transonic points, where flow turns from subsonic to supersonic (blue dots in Fig.~\ref{fig:schematics}). Only proper boundary condition can avoid hydraulic jumps across these singular points. One way to solve the problem is to do a binary search for the boundary condition so that unphysical jumps and backflow are avoided \citep{Ingersoll-Summers-Schlipf-1985:supersonic}. However, thanks to the conservation of mass (see below), and thus condensation releases latent heating but doesn't cause mass loss in the mineral vapor flow), energy and Bernoulli function on day-side, we found a way to avoid the binary search and instead directly solve for the escape flux and boundary condition. But for the night-side escape flow and the day-side to night-side transport flow, binary search seems to be necessary due to the break down of mass conservation (see below). We leave technical details to the appendix.

On the day-side, we assume that dusts, once formed, would be re-vaporized almost immediately by the stellar radiation. This is justified for sodium or SiO dominant atmosphere, because, as shown in appendix~\ref{sec:appendix-vertical-escape-day}, even a $10~\mu$m sodium droplet will completely vaporize in the matter of a few seconds or less, and a SiO particle will vaporize within a few minutes. As a result, condensation releases latent heating but doesn't cause mass loss in the mineral vapor flow.
Nevertheless, for the night-side escape flow, where no stellar radiation is received, and for the transport flow, where mass exchange keeps happening between the atmosphere and the planet's surface, we instead assume dusts stop interacting with the remaining vapor flow, once formed.

In principle, the model described above can be applied to any arbitrary chemical components. Here, we consider a sodium dominant escape flow as in \citet{Mura-Wurz-Schneider-et-al-2011:comet}, motivated by the ubiquity of sodium tail within the solar system (sodium tail has been observed around the moon \citep{Matta-Smith-Baumgardner-et-al-2009:sodium}, Mercury \citep{Potter-Killen-Morgan-2002:sodium} and also Comet Hale-Bopp \citep{Cremonese-Huebner-Rauer-et-al-2002:neutral}). It is possible that sodium is also major composition of escape flow on planets beyond the solar system. Actually, as long as sodium has not been exhausted from the magma ocean, its high volatility would make it, by far, the dominant component of the escape flow \citep{Schaefer-Fegley-2009:chemistry}. However, since sodium only accounts for 0.29\% of the total mantle mass with bulk silicate earth composition \citep{Schaefer-Fegley-2009:chemistry}, sodium could be exhausted in the early stage of a planet's lifetime depending on the efficiency of mantle-surface exchange \citep{Kite-Fegley-Schaefer-et-al-2016:atmosphere}. We therefore repeat the calculation assuming a SiO-dominant escape flow; the results are qualitatively similar.

In all calculations, we use planetary orbital parameters taken from KIC-1255b \citep{Brogi-Keller-Juan-et-al-2012:evidence, Lieshout-Rappaport-2018:disintegrating}, fix the planetary density to be the same as Earth, and investigate how the escape flow properties vary with substellar surface temperature $T_0$ and planet mass $M_p$. The parameters used in this study are summarized in Table.~1 in the appendix.

\section{Day-side and night-side escapes on KIC-1255b as an example}

 \begin{figure*}[htpb!]
  \centering \includegraphics[width=0.75\textwidth]{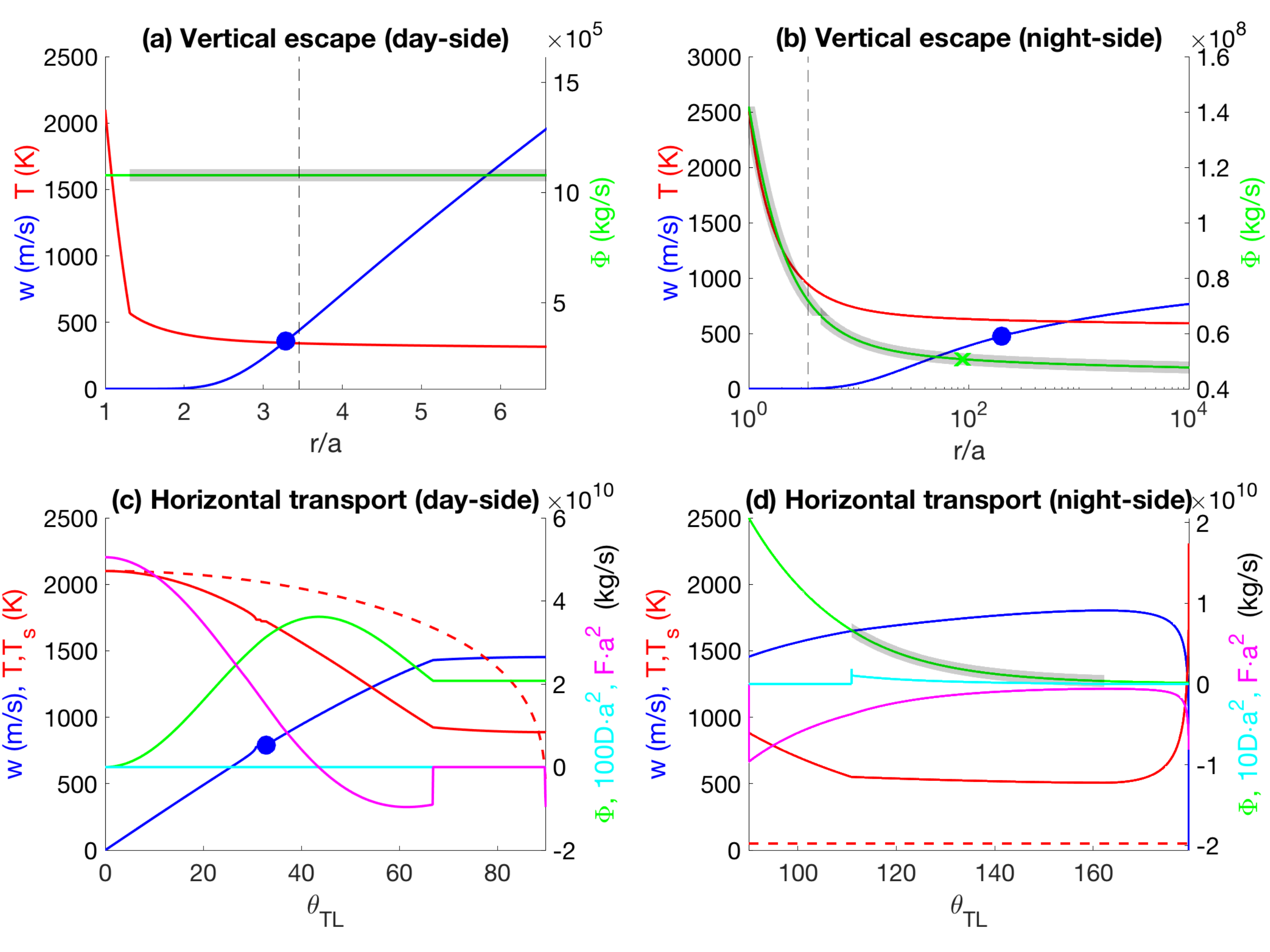}
  \caption{Properties of escape flow and transport flow for $T_0=2100~K, M_p=0.03M_{\mathrm{earth}}$. (a) and (b) show the radial dependence of temperature $T$ (red), speed $w$ (blue) and mass flux $\Phi$ (green) for day-side escape and night-side escape, respectively. The vertical thin dashed line denotes the Roche lobe radius. Panel (a) shows the profiles for the day-side escape flow. Transonic point is marked by a blue dot on the speed curve. Wherever condensation occurs, a gray shading is overlaid on top of the mass flux curve. For the purpose of representation, we consider the whole magma ocean as one band with surface temperature set to substellar point, rather than split it into 10 bands as we do for real escape calculations shown in Fig.~\ref{fig:escape-summary} and Fig.~\ref{fig:escape-summary-sio}. The green cross in panel (b) marks the mass flux that can finally escape the gravity of the planet; matters that condensed before this point will fall back to the surface. (c) and (d) show the properties of the horizontal transport flow from the day-side to the night-side. $\theta_{\mathrm{TL}}$ is the tidally locked latitude, angular distance from the substellar point. Red dashed curves show the surface temperature profiles, cyan curves and magenta curves show the condensation rate $D$ and surface exchange flux $F$ multiplied by $a_p^2$. The rest curves are defined the same way as in (a) and (b).}
  \label{fig:escape-example}
\end{figure*}

 \begin{figure*}[htpb!]
  \centering \includegraphics[width=0.75\textwidth]{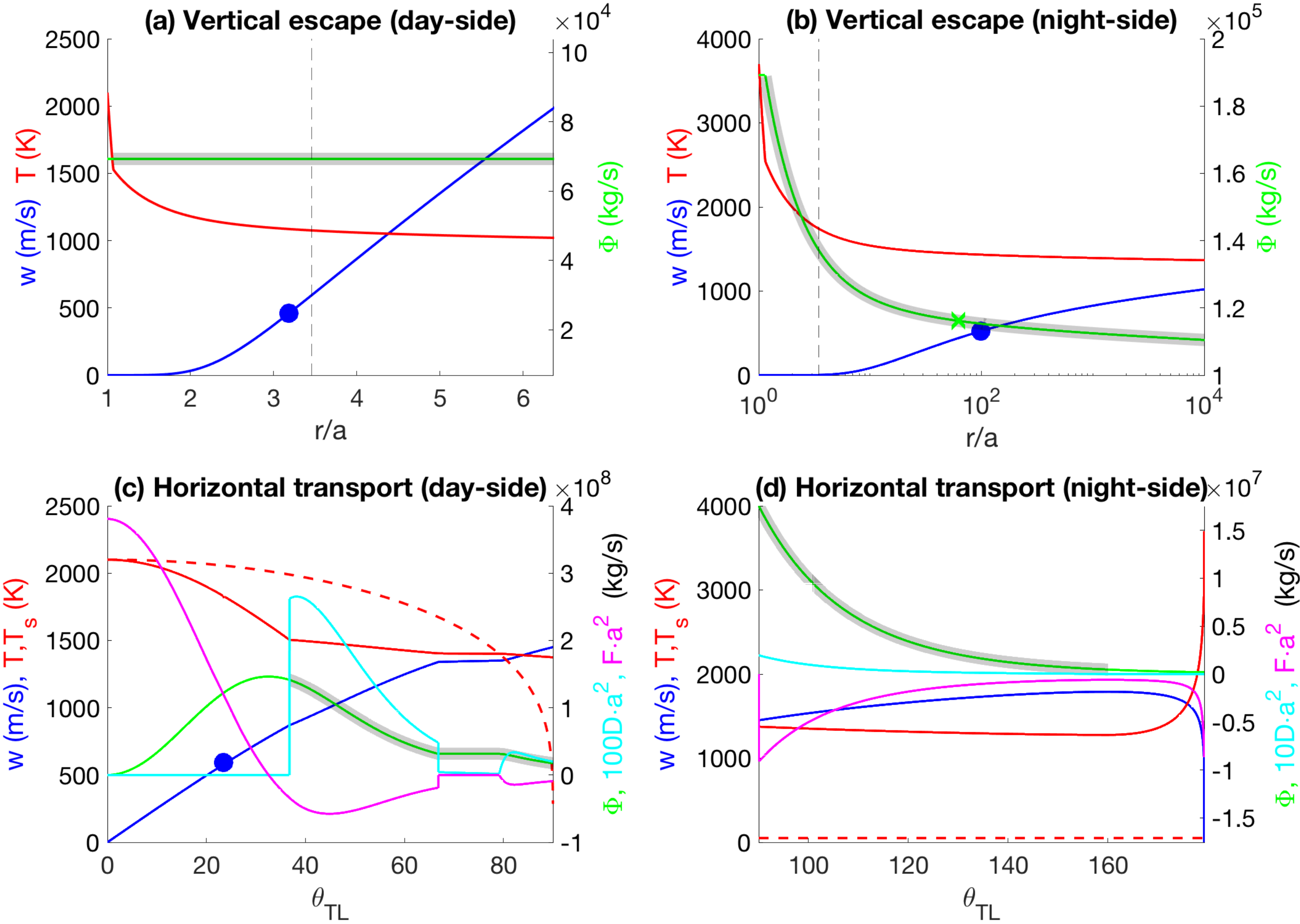}
  \caption{Same as Fig.~\ref{fig:escape-example} except for SiO dominant atmosphere.}
  \label{fig:escape-example-sio}
\end{figure*}

An example of the escape process is shown in Fig.~\ref{fig:escape-example}. Surface temperature is set to 2100~K and planetary mass is set to $0.03$ Earth mass.

In the day-side escape pathway (Fig.~\ref{fig:escape-example}a), the flow is initially undersaturated, due to the dilution of other chemical components. Vapor cools quickly upward, as internal energy is converted to kinetic energy and gravity potential energy. At around $1.3$ planetary radii, flow becomes saturated. Beyond the saturation level, latent heat release due to condensation becomes the main energy source, preventing further drop of temperature. However, the particles formed due to condensation almost immediately re-evaporate due to the strong stellar radiation (the lifetime of sodium droplet is below a second, see appendix~\ref{sec:appendix-vertical-escape-day} for details). The continuous condensation and re-evaporation assure no loss in mass in the day-side escape flow $\Phi$ (i.e., $\Phi$ is a constant). Although mass flux is not changed, this recycling keeps energizing the flow until it escapes the gravity field of the planet. The flow turns supersonic at around $3.2$ planetary radii (see blue dot in Fig.~\ref{fig:escape-example}a), right before the Roche lobe (the thin dashed line in Fig.~\ref{fig:escape-example}a).

The night-side escape pathway takes three steps: flow first gathers mass from the day-side while accelerating toward the night-side (Fig.~\ref{fig:escape-example}c); subsequently, its speed drops and pressure rises, as flow converges near the anti-stellar point (Fig.~\ref{fig:escape-example}d); finally, the high pressure and air temperature at the surface drives escape flow away the planet (Fig.~\ref{fig:escape-example}b).

Near the substellar point, sodium vaporizes from the magma ocean (positive surface flux $F$ shown by magenta curve in Fig.~\ref{fig:escape-example}c), because the vapor pressure is lower than that required by chemical equilibrium with the magma ocean (about 50\% lower in this example). Sodium gas enters the atmosphere with the same temperature as the magma ocean beneath (atmospheric temperature (red solid curve) overlaps with the surface temperature (red dashed curve)). Away from the substellar point, surface cools and the corresponding equilibrium pressure drops exponentially along with surface temperature, creating a pressure gradient force accelerating the transport flow toward the night-side (see wind speed shown in blue solid curve). In this process, the internal energy of the flow is converted to kinetic energy -- as wind accelerates, the atmosphere becomes cooler than the surface. Around 33$^\circ$, flow turns supersonic (marked by a blue dot). Beyond this transonic point, no information can be transported upstream to affect the mass flux and flow properties before that point. Surface temperature and the associated equilibrium pressure keep dropping toward the night-side. At around 43$^{\circ}$, the equilibrium pressure drops below the flow pressure, and sodium starts to infuse back into the magma ocean ($F$ turns negative). At around 67$^{\circ}$, the planet's surface solidifies, and therefore can no longer provide a source of mineral vapor (sodium, in our model). That means the transported mass flux $\Phi$ would at most remain the same, if not be attenuated. While the surface is warmer than the flow above it (this is true for most of the day-side hemisphere), no infusion would occur and $\Phi$ remains unchanged.

When flow reaches the night-side (Fig.~\ref{fig:escape-example}d), where the surface temperature is always lower than the flow, vapor starts to infuse into the surface, as represented by a negative $F$. The attenuation of $\Phi$ leads to a pressure drop, which acts to accelerate and cool the flow. Along the way, supersaturation and condensation would occur (condensation is shown in a cyan curve, and the saturated regions are marked with a gray shading). Pressure drop keeps accelerating the flow until geometric convergence near the antistellar point causes mass accumulation. The mass accumulation give rise to a surface high pressure, which slows down the flow and turns it upward\footnote{We let the flow turn once its speed drops over 50\% from the peak. Results are not sensitive to this arbitrary choice. As shown later, a reasonably good analytical estimation can be obtained without using this information.}. Mass and energy are assumed to be conserved during this turning process.

The vertical escape process is shown in Fig.~\ref{fig:escape-example}b. The flow density and pressure starts high, accelerating the upward escape flow. Flow also become saturated soon after departing from surface, if not already saturated at the surface, as its potential and kinetic energy increases at the expense of reducing its internal energy. Once saturated, condensation prevents further temperature drop in the escape flow (marked by gray shading in Fig.~\ref{fig:escape-example}b), and meanwhile the mass flux reduces. At the cost of losing mass flux, the rest of the flow achieves enough energy to escape from the planet's gravity field, inheriting the latent heating released by condensation. Around 200 planetary radii, flow turns supersonic, beyond which all mass has to escape. Assuming no momentum exchange between dust and the rest of gas, dusts can escape the planet's gravity if and only if the total mechanical energy (kinetic energy plus potential energy) is greater than zero. In this example, this occurs around 90 planetary radii, marked by a green cross in Fig.~\ref{fig:escape-example}b.

The same calculation assuming a SiO-dominant atmosphere yields a similar escape process (see Fig.~\ref{fig:escape-example-sio}), except two main differences. First, since the SiO equilibrium pressure at the surface is closer to saturation compared to sodium (due to the higher percentage of SiO in the magma), SiO-dominant atmosphere starts to condense even in the day-side of the horizontal transport flow (Fig.~\ref{fig:escape-example-sio}c). Second, the escape flux is 2-3 orders of magnitude smaller than that we obtained assuming a sodium-dominant atmosphere, which is expected because the equilibrium pressure of SiO is much lower than that of sodium (this will be discussed by the end of next section).

\section{Day-side escape versus night-side escape for generic exoplanets}
\label{sec:escape-generic-planets}
\begin{figure*}[htpb!]
  \centering
\includegraphics[width=0.65\textwidth]{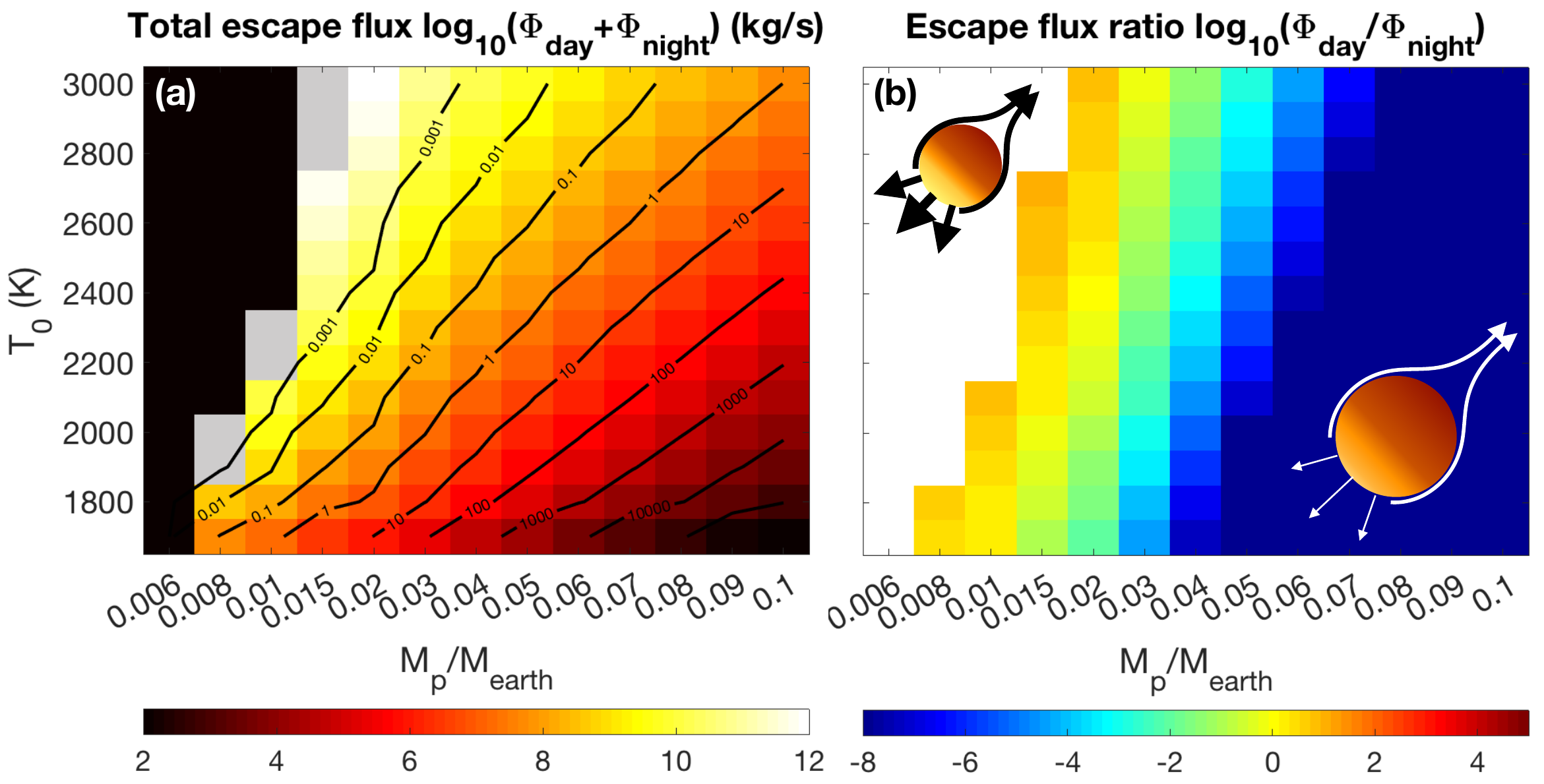}
  \caption{Summary of escape properties for various $(T_0,\ M_p)$ combinations. Shown from left to right are (a) logarithm of total escape flux in kg/s, (b) logarithm of the ratio between day-side escape and night-side escape. In panel (a), the parameter combinations where atmosphere escape would consume at least the same order of magnitude of energy as the planetary thermal emission is marked by a gray shading; and the parameter combinations leading to ``run-away'' escape is marked by a black shading on the upper-left corner. The e-folding time for the sodium reservoir ($0.29\%$ of total planetary mass) to be exhausted is shown by contours in the units of Myr. Different flow field features are illustrated in panel (b): in the small and hot limit, escape is mostly contributed by the direct day-side escape, while the night-side escape dominates in the large and cool limit. }
  \label{fig:escape-summary}
\end{figure*}

Using this idealized model, we can calculate the escape flux from the day-side and night-side pathways for generic disintegrating exoplanets. The total escape flux is shown in Fig.~\ref{fig:escape-summary}(a) as a function of peak surface temperature and the planetary mass. As expected, the total escape increases with temperature and decreases with the planet's mass. Toward the limit of hot and small planets, escape flux becomes so strong that the energy required to maintain it has the same order of magnitude as the total planetary thermal emission (gray shading). For planets in this regime, insolation has to be significantly stronger than what the planet's surface can emit, in order to maintain the escape flow. Black shading marks the regime of the ``run-away'' escape state, as found by \citet{Lehmer-Catling-Zahnle-2017:longevity} and \citet{Arnscheidt-Wordsworth-Ding-2019:atmospheric}. In reality, the infinite escape flow is not achievable, because escape flux would require so much energy that the surface temperature would fall below the ``run-away'' threshold.

\begin{figure*}
  \centering
\includegraphics[width=0.65\textwidth]{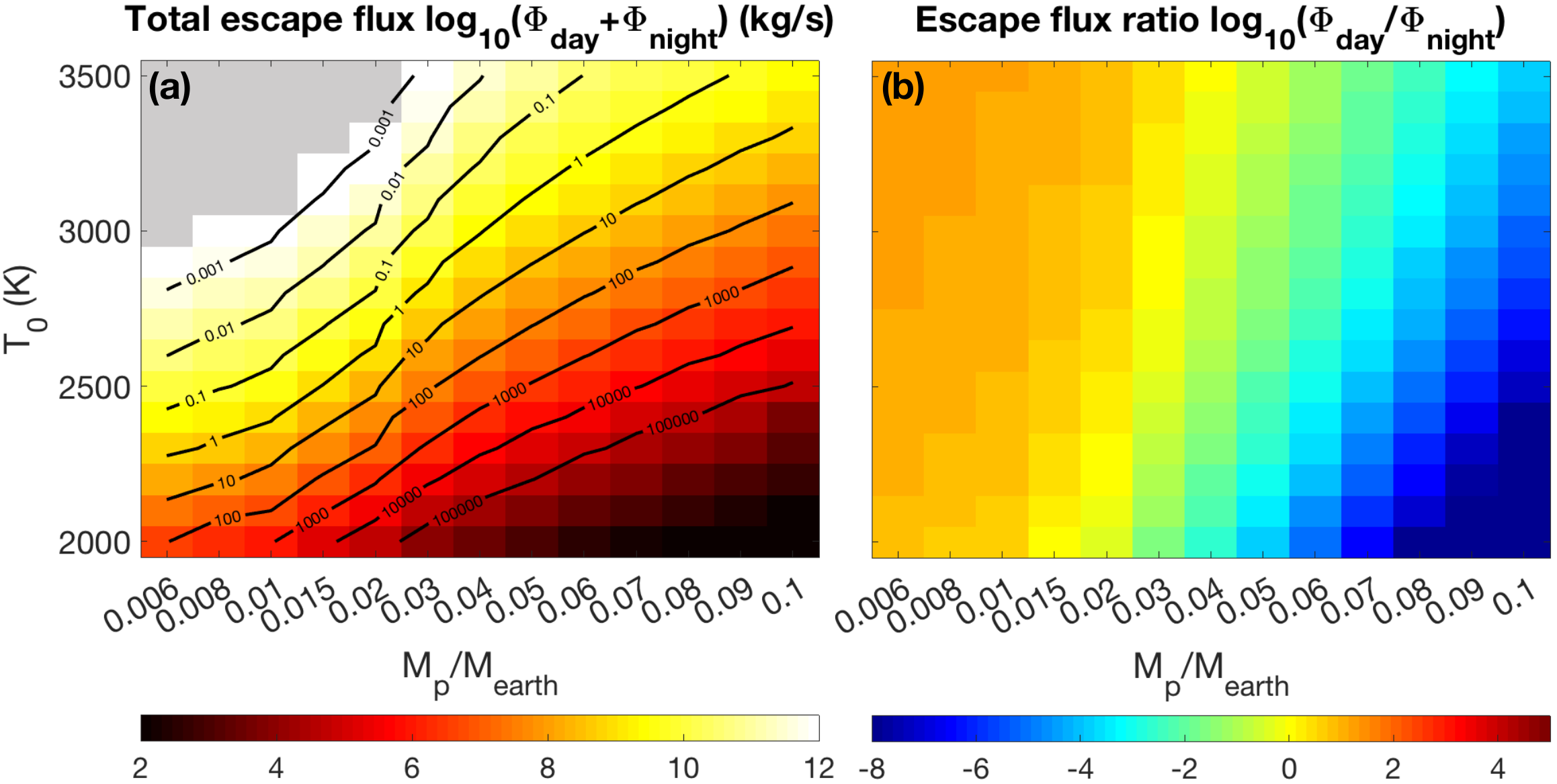}
  \caption{Same as Fig.~\ref{fig:escape-summary}(a,b) except the atmosphere is SiO instead of Na.}
  \label{fig:escape-summary-sio}
\end{figure*}

Atmospheric escape would gradually exhaust the sodium reservoir on the planet; the reduced sodium concentration would in turn weaken the escape flow. The e-folding time of the sodium reservoir can be estimated as
\begin{equation}
  \label{eq:exhause-e-fold}
  \tau_{\mathrm{Na}}=c_{\mathrm{Na}}M_p/\Phi,
\end{equation}
where $c_{\mathrm{Na}}=0.29\%$ is the initial mass concentration of sodium for Bulk Silicate Earth according to \citet{Schaefer-Lodders-Fegley-2012:vaporization}, $M_p$ is the mass of the planet, and $\Phi$ denotes the total escape rate. In this estimation, $\Phi$ is assumed to vary in proportion with the sodium concentration. We show $\tau_{\mathrm{Na}}$ (units: Myr) in Fig.~\ref{fig:escape-summary}a using black contours. The isolines of $\tau_{\mathrm{Na}}$ are mostly aligned with that of the escape flux $\Phi$ (shading), indicating that escape rate is the key factor in determining the ``lifetime'' of the sodium reservoir.
According to the estimation by \citet{Perez-Becker-Chiang-2013:catastrophic} based on the transit depth, the escape flux should be around $10^7$-$10^8$~kg/s for KIC-1255b, KOI-2700b and K2-22b. This corresponds to an e-folding time of around 0.1-1~Myr for the sodium reservoir, according to Fig.~\ref{fig:escape-summary}a. After that, SiO and Mg, whose initial pressure contributions are 1-5 orders of magnitude smaller than sodium's \citep{Schaefer-Fegley-2009:chemistry, Miguel-Kaltenegger-Fegley-et-al-2011:compositions}, would replace sodium becoming the main component of the atmosphere \citep{Kite-Fegley-Schaefer-et-al-2016:atmosphere}. A similar calculation can then be carried out for the SiO/Mg dominant atmosphere (we repeat the calculation for SiO dominant atmosphere in Fig.~\ref{fig:escape-summary-sio}), and so on and so forth. 

Although both day-side and night-side escape decrease as temperature drops or planetary size increases, the day-side escape rate drops at a much faster rate, giving rise to a transition from day-side dominant escape to night-side dominant escape, as shown in Fig.~\ref{fig:escape-summary}b. Physically, this is because the day-side escape flow is driven by the difference between the pressure gradient force and the gravity, while the day-to-night transport flow is driven by the pressure gradient force as a whole. The strong gravity on a relatively large planet would directly suppress the day-side escape flow, but won't significantly affect the horizontal transport flow. The flow transported to the night-side has to either escape from the planet or condense and fall onto the surface. Gravity will affect the percentage of mass flux that can eventually escape, but this factor is order 1\footnote{As shown in section~\ref{sec:analytical-night}, the escape percentage is $exp(\Phi(a)/L)$, where $\Phi(a)$ is the gravity potential at the surface with reference set at infinity, $L$ is the sublimation enthalpy.). In the relevant parameter regime, $\Phi(a)$ is smaller or at most comparable with $L$, otherwise the escape flow would be very weak.}. As a result, the escape flow from the night-side dominates that from the day-side when the planet is relatively large.

FUV flux from the star could trigger additional day-side escape. With a FUV flux of $0.45$~W/m$^2$ (450~erg/s/cm$^2$), energy limit yields an escape rate of $6\times 10^5$~kg/s (details are described in appendix~\ref{sec:appendix-vertical-escape-day}), and this is a dominant contributor only in the regime marked by dark red or black in Fig.~\ref{fig:escape-summary}a.

By conserving angular momentum, day-side escape flow would lead in front of the planet in orbit, while night-side escape flow would lag behind. The latter provides a natural explanation for the trailing tails of the disintegrating exoplanets \citep{Brogi-Keller-Juan-et-al-2012:evidence, Rappaport-Barclay-DeVore-et-al-2014:koi, Rappaport-Levine-Chiang-et-al-2012:possible, Sanchis-Ojeda-Rappaport-Palle-et-al-2015:the, Budaj-Kocifaj-Salmeron-et-al-2015:tables}, which used to be explained by dusts carried upward by the escaping flow and pushed backward by the stellar radiation pressure \citep{Rappaport-Barclay-DeVore-et-al-2014:koi, Sanchis-Ojeda-Rappaport-Palle-et-al-2015:the, Lieshout-Min-Dominik-2014:dusty}.

In order for dusts to significantly lag behind the planet in orbit, the radiation pressure needs to be strong and the lifetime of the dusts needs to be comparable with the orbital period. Otherwise, the constant mass exchange between condensed phase and gas phase guarantees that gas and dusts travel along the same trajectory and the mass-weighted bulk radiation pressure is likely to be weak diluted by the large amount gas phase materials. This, in turn, requires dusts to be made of refractory compositions such as corundum and fayalite \citep{Lieshout-Min-Dominik-2014:dusty}. However, the evaporation of refractory materials from the magma ocean is likely to be inhibited in the first place (thermodynamic calculations suggest that the atmosphere composition above a magma ocean with bulk silicate earth composition is mainly made of much more volatile species such as Na, SiO, Mg and Fe, even at a temperature as low as 1000K \citet{Miguel-Kaltenegger-Fegley-et-al-2011:compositions, Schaefer-Fegley-2009:chemistry, Schaefer-Lodders-Fegley-2012:vaporization}). Instead, these refractory composition seems more likely to be produced by chemical reactions as flow cools, although the possibility remains that strong escape flow can knock off a significant amount of refractory materials from the surface and carry them upward.

An alternative physical picture is that condensation of relatively volatile species as shown in Fig.~\ref{fig:escape-example}(a,b) keeps occurring until the flow is far away from the planet ($\sim 10$ planetary radii), and this continuous particle source may help explain the length of the tails \cite{Brogi-Keller-Juan-et-al-2012:evidence, Rappaport-Barclay-DeVore-et-al-2014:koi, Sanchis-Ojeda-Rappaport-Palle-et-al-2015:the, Budaj-Kocifaj-Salmeron-et-al-2015:tables}. 
However, condensation occurs mostly within 10 planetary radii, and this is by far not enough to explain the observed tail whose length is comparable to the stellar radius \citep{Brogi-Keller-Juan-et-al-2012:evidence, Rappaport-Barclay-DeVore-et-al-2014:koi, Rappaport-Levine-Chiang-et-al-2012:possible, Sanchis-Ojeda-Rappaport-Palle-et-al-2015:the, Budaj-Kocifaj-Salmeron-et-al-2015:tables}. Energetically, cooling and condensation would only occur when flow is doing work against gravity or accelerating. After 10 planetary radii, the gravity of the planet would be too weak to induce any condensation. Condensation beyond that distance is possible only if the flow is accelerating rapidly due to the pressure drop induced by the expansion of cross section, or the flow continues to do work against stellar gravity. 

Interestingly, a lot of the parameter combinations we explored here yield a night-side escape that is even stronger than the day-side escape. Besides this, a day-side dominant escape only occurs on planets that are hot, small and hence short-lived. This further reduces the probability of observing such planets. Whether leading tail or trailing tail will be more common is a more complex question that requires consideration of the dusts' optical properties and the trajectory of the leading/trailing tails (more specifically, how elliptical the trajectory would be), and is out of the scope of current work. However, the ratio between the day-side and night-side escape flux may provide a rough estimation for the relative significance of the leading tail and trailing tail. The observations of tail properties may be able to put constraints on the mass and radius of these disintegrating exoplanets, which are poorly constrained because of their small sizes and thick dusty envelopes \citep{Lieshout-Rappaport-2018:disintegrating}.

We also repeat the same calculation for SiO-dominant atmosphere and obtain qualitatively similar results (Fig.~\ref{fig:escape-summary-sio}). Compared to the sodium-dominant escape (Fig.~\ref{fig:escape-summary}), SiO escapes at a much slower rate. At a given temperature and planetary mass, SiO escape is more difficult to be detected. 

However, in reality, the atmosphere should be a mixture of volatile components (such as sodium and oxygen), less volatile components (such as SiO, Fe, Mg) and a trace amount of refractory components (such as fayalite, pyroxene and corundum). The dominant component is not necessarily the one that condenses, reacts and thus provides energy. Actually, volatile species tend to take a large portion in mass, but they less tempted to condense. If we assume that the mixing ratio of different compositions stays fixed during the escape and transport processes\footnote{This is true for the day-side escape flow as the dusts tend to be melted soon after they form under the fierce stellar radiation. Our model for the night-side escape flow assumes that the night-side flow remains in the shade of the planet, and thus, no evaporation would happen. But in reality, the night-side flow would lag behind the planet in orbit as it heads away from the star, according to angular momentum conservation. Melting would also occur after flow leaves the planetary shade.}, then the total pressure of the mixture is likely to be much greater than the pressure of a pure SiO atmosphere. This greater pressure would in turn enhances escape flow. This calculation would requires one to know all the potential condensible/reactive matters in the flow and how the saturated/equilibrium pressure depends on temperature, and thus is beyond the scope of our work.

Finally, we note that the time it takes to exhaust SiO/sodium is not long (far below 1~Gyr) according to the calculation. This doesn't mean that we wouldn't detect disintegrating planets at all, although it does mean that only a small portion of existing planets will be detected. The likelihood to detect a disintegrating planet at a certain temperature and initial mass and the likelihood to detect a planet with trailing tail versus leading tail is yet to be explored.

\section{Analytical approximation for escape mass flux}
\label{sec:analytical}

In this section, we make further simplifications to the escape model to obtain analytical approximations for the escape mass flux from the day-side and the night-side.

\subsection{For day-side}
\label{sec:analytical-day}
Using conservation laws and transonic conditions, we have already been able to semi-analytically solve for the day-side escape flux from a set of algebratic equations without integrating or iterating the 1D hydrodynamic escape model as done in \citet{Lehmer-Catling-Zahnle-2017:longevity}. Details are in section~\ref{sec:appendix-vertical-escape-day}. Here, we make more simplifications to get an analytical approximation. 

The first simplification we make is to ignore the surface temperature variations across magma ocean and assume the day-side escape flow starts with the mean surface temperature $\bar{T}$ and the corresponding chemical equilibrium pressure $\bar{P}$ within magma ocean 
\begin{eqnarray}
  \label{eq:day-predict-Pmagma}
  &~&\bar{P}\sim P_{\mathrm{chem}}\left(\bar{T}\right)=A\exp(-B/\bar{T})\\ &~&\bar{T}=\left.\int_0^{\theta_b}T_s(\theta')\cos\theta'd\theta'\right/\int_0^{\theta_b}\cos\theta'd\theta',\label{eq:day-predict-Tmagma}
\end{eqnarray}
where $T_s$ is the surface temperature set by radiative equilibrium, $\theta_b$ is the latitude where the surface solidifies, $T_s(\theta_b)=1673$K.
$A=10^{9.6}$~Pa$=10^{10.6}$~Ba and $B=38000$K \citep{Castan-Menou-2011:atmospheres}.

The escape flow is initially unsaturated, due to the dilution of other components in the magma ocean (Henry's law) and will follow the dry adiabatic profile, $TP^{-\kappa}=$const., where $\kappa=R/C_p$, $C_p$ is the vapor's heat capacity at constant pressure, and $R$ is the gas constant. Saturation would occur at $T_{\mathrm{sat}}$. This condensation temperature can be determined by finding the intersection between the dry adiabat and the Clausius-Clapeyron relation. 
\begin{eqnarray}
  \label{eq:day-predict-Tcond}
T_{\mathrm{sat}}P_{\mathrm{sat}}(T_{\mathrm{sat}})^{-\kappa}=\bar{T}P_{\mathrm{chem}}(\bar{T})^{-\kappa} 
\end{eqnarray}
In above equation, $P_{\mathrm{sat}}(T)=A_{\mathrm{sat}}\exp\left(-\frac{B_{\mathrm{sat}}}{T}\right)$ is the saturated vapor pressure at $T$. $A_{\mathrm{sat}}=10^{9.54}$~Pa$=10^{10.54}$~Ba and $B_{\mathrm{sat}}=12070.4$K for sodium \citep{Bowles-Rosenblum-1965:vapor}. This saturation would occur around $r_{\mathrm{sat}}$ from the center of the planet, where the gain of gravity potential energy matches the loss of internal energy since the flow departures from the surface (we ignore the change of kinetic energy since they are small according to Fig.~\ref{fig:escape-example}a).
\begin{eqnarray}
  \label{eq:day-predict-rsat}
  \Psi(r_{\mathrm{sat}})-\Psi(a)&=&C_p(\bar{T}-T_{\mathrm{sat}})
\end{eqnarray}
Here, $\Psi(r)$ denotes the gravity potential at a distance $r$ from the center of the planet toward the host star.
   \begin{equation}
    \label{eq:day-predict-Psi}
 \Psi(r)=-\frac{ga^2}{r}-\frac{3}{2}\Omega^2r^2,   
\end{equation}
where $g,a$ are the surface gravity and radius of the planet, and $\Omega$ is the orbital angular speed. The second term comes from the tidal force.

With $P_{\mathrm{sat}}$ in such a form, one can find $w^2/2+RB_{\mathrm{sat}}\ln(T)+\Psi(r)$ conserved as long as the flow is saturated (see section~\ref{sec:appendix-vertical-escape-day} for derivation). Here, $w$ denotes vertical flow speed, and $T$ denotes flow temperature. This conserved quantity can link the flow state at $r_{c}$, the transonic point, and $r_{\mathrm{sat}}$, the point where flow just turn saturated,
\begin{equation}
  \label{eq:day-predict-conserve-bernoulli}
  \frac{w_c^2}{2}+RB_{\mathrm{sat}}\ln(T_c)+\Psi(r_c)=\frac{w_{\mathrm{sat}}^2}{2}+RB_{\mathrm{sat}}\ln(T_{\mathrm{sat}})+\Psi(r_{\mathrm{sat}}),
\end{equation}
where $w_c$ and $T_c$ denote the flow speed and temperature at $r_c$.
In the above equation, $w_{\mathrm{sat}}$ is negligible compared to $w_c$ (according to Fig.~\ref{fig:escape-summary}a). $w_c$ can be expressed with $T_c$ using transonic conditions \citep{Lehmer-Catling-Zahnle-2017:longevity, Pierrehumbert-2010:principles},
 \begin{equation}
   \label{eq:day-predict-transonic-cond-moist}
   w_c^2=\frac{RT_cB_{\mathrm{sat}}}{B_{\mathrm{sat}}-T_c}.
 \end{equation}
 Since $T_c\sim 10^3~K$, $w_c$ is around $500$~m/s, and thus the term $w_c^2/2$ in Eq.~(\ref{eq:day-predict-conserve-bernoulli}) is at least 2-3 orders of magnitude smaller than $RB_{\mathrm{sat}}\ln(T_c)$. Dropping the $w_c$ and $w_{\mathrm{sat}}$ terms in Eq.~(\ref{eq:day-predict-conserve-bernoulli}) and assuming that $\Psi(r_c)$ is close to the maximum gravity geopotential between the planet and the star, which is achieved at the Roche lobe $r_{\mathrm{Roche}}=(ga^2/3\Omega^2)^{1/3}$, we obtain an approximation for $T_c$ with the help of Eq.~(\ref{eq:day-predict-rsat}),
  \begin{equation}
   \label{eq:day-predict-Tc} T_c=T_{\mathrm{sat}}\exp\left(\frac{\Psi(a)-\Psi(r_{\mathrm{Roche}})+C_p(\bar{T}-T_{\mathrm{sat}})}{RB_{\mathrm{sat}}}\right).
 \end{equation}
 
 Utilizing the transonic conditions (Eq.~\ref{eq:day-predict-transonic-cond-moist}) again, we get an approximation for the day-side escape flux
 \begin{eqnarray}
\Phi_{\mathrm{day}}&\approx&w_c\frac{P_{\mathrm{sat}}(T_c)}{RT_c}\cdot (2\pi r_c^2(1-\cos\theta_b))\nonumber\\
                   &=& \frac{\pi A_{\mathrm{sat}} g^2a^4(1-\cos\theta_b)}{2RT_c} \left(\frac{B_{\mathrm{sat}}-T_c}{RT_cB_{\mathrm{sat}}}\right)^{\frac{3}{2}}\exp\left(-\frac{B_{\mathrm{sat}}}{T_c}\right)\nonumber\\
                   &\approx& \frac{\pi A_{\mathrm{sat}} g^2a^4(1-\cos\theta_b)}{2(RT_c)^{5/2}}\exp\left(-\frac{B_{\mathrm{sat}}}{T_c}\right)
                       \label{eq:day-predict-Phi}
 \end{eqnarray}
 Again, $\theta_b$ is the latitude where the surface solidifies.

\subsection{For night-side}
\label{sec:analytical-night}
The night-side escape flow is determined by how much air mass is transported from day-side to night-side and how much of this mass flux can eventually escape (demonstrated by Fig.~\ref{fig:escape-example}c,d,b). Exact analytical solution cannot be achieved because the conservation laws are broken by condensation and by mass, momentum and energy exchanges between transport flow and the magma ocean underneath. However, an analytical approximation is possible. 

In the first stage, mineral vapor is gathered from magma ocean and transported to the terminator. To estimate the cross-terminator mass flux, we need to know the flow speed and column mass (pressure divide by gravity acceleration $g$, assuming hydrostatic balance). As shown in Fig.~\ref{fig:escape-example}c, the solidified surface on the day-side almost plays no role in changing transport flux (green curve flattens beyond magma ocean). Within magma ocean, the flow pressure is relaxed toward the chemical equilibrium pressure, which is set by the magma ocean surface temperature. Again, we assume the magma ocean has a uniform temperature $\bar{T}$ (Eq.~\ref{eq:day-predict-Tmagma}), which corresponds to an equilibrium pressure of $\bar{P}$ (Eq.~\ref{eq:day-predict-Pmagma}).

\begin{figure*}
  \centering
\includegraphics[width=0.75\textwidth]{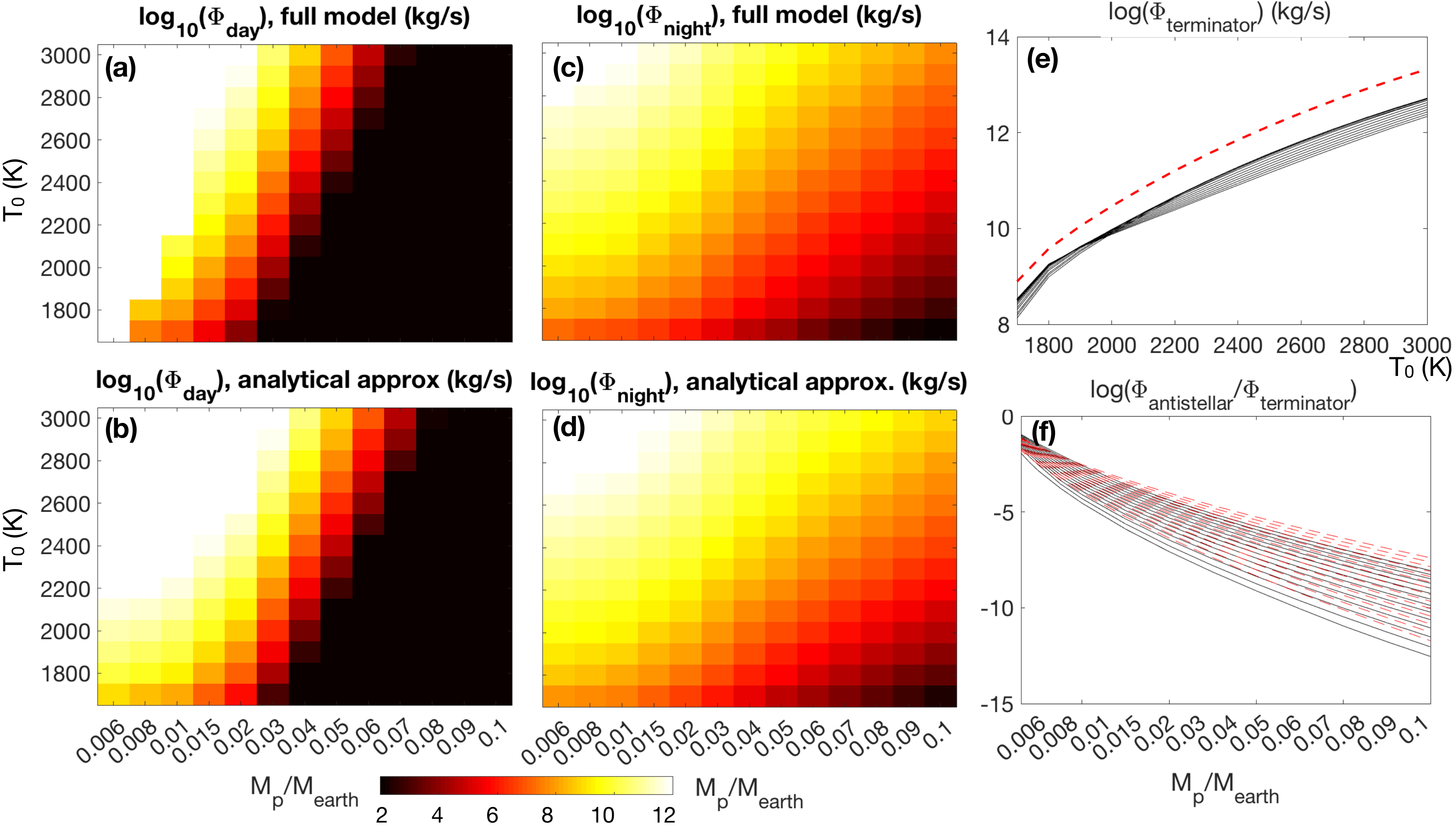}
  \caption{Analytical estimation for the sodium escape from day-side and night-side. Panel (a,b) show day-side escape flux for various $\{T_0,M_p\}$ from full model and from analytical estimation (Eq.~\ref{eq:day-predict-Phi}), respectively. Panel (c,d) are the same as panel (a,b), but for night-side. Eq.~(\ref{eq:night-predict-Phi}) gives the analytical estimation for night-side escape rate. Panel (e) shows the mass flux across terminator as a function of substellar temperature $T_0$. Black curves are mass flux diagnosed from the full transport model, and red dashed curves are the analytical estimation given by Eq.~(\ref{eq:night-predict-Phiterminator}). Each black curve corresponds to a different planet mass $M_p$. Panel (f) is similar to (e) but shows the mass flux decay on night-side as a function of $M_p$. Curves from top to bottom correspond to decreasing $T_0$. }
  \label{fig:daynight-escape-predict}
\end{figure*}

Flow usually turns supersonic within magma ocean (Fig.~\ref{fig:escape-example}c). We here take the sonic speed at $\bar{T}$ as an estimation of flow speed,
\begin{equation}
  \label{eq:night-predict-Vflow}
  V_{\mathrm{flow}}\sim\sqrt{R\bar{T}},
\end{equation}
where $R$ is the specific gas constant. Putting Eq.~(\ref{eq:day-predict-Pmagma}) and Eq.~(\ref{eq:night-predict-Vflow}) together yields an estimation of the cross-terminator mass flux
\begin{eqnarray}
  \label{eq:night-predict-Phiterminator} \Phi_{\mathrm{terminator}}&=&V_{\mathrm{flow}}\bar{P}/g\cdot (2\pi a\sin\theta_b)\nonumber\\
  &\sim& 2\pi A\exp\left(\frac{-B}{\bar{T}}\right) \sqrt{R\bar{T}}\sin\theta_b(\frac{a}{g}).
\end{eqnarray}
This estimation suggests that $\Phi_{\mathrm{terminator}}$ would increase with $T_0$, and that $\Phi_{\mathrm{terminator}}$ is not sensitive to the planet size: the only factor directly related to planet size, $a/g$, is a fixed constant when planetary density is fixed. The dependence of $\Phi_{\mathrm{terminator}}$ on $T_0$ is shown in Fig.~\ref{fig:daynight-escape-predict}(e). Black curves are $\Phi_{\mathrm{terminator}}$ diagnosed from the 1D transport model. Each curve corresponds to a different planet mass, and the collapse between different curves validates the insensitivity to planet size. Red dashed curve shows the analytical estimation in Eq.~(\ref{eq:night-predict-Phiterminator}). The variation of $\Phi_{\mathrm{terminator}}$ with $T_0$ is well captured except a slight overestimation.

In the second stage, flow infuses back to surface while passing through the cold night-side. Assuming every molecule colliding with the surface get trapped, the attenuation of mass flux $\Phi$ on the night-side can be calculated
\begin{equation}
  \frac{1}{a\sin\theta} \frac{d\Phi}{d\theta}=-\frac{P}{\sqrt{2\pi RT}}(2\pi a) \label{eq:night-predict-massloss1}
\end{equation}
Substituting the definition $\Phi=(VP/g)(2\pi a\sin\theta)$, we get
\begin{equation}
  \label{eq:night-predict-massloss}
  \frac{d\ln\Phi}{d\theta}=-\frac{ga}{V\sqrt{2\pi RT}}
\end{equation}
Assuming almost all internal energy has been converted to kinetic energy before flow enters the night-side would allow us to substitute $V$ with $\sqrt{2C_p\bar{T}}$, where $C_p$ is the vapor's heat capacity at constant pressure. Condensation is likely to occur on the night-side, and then the flow temperature is fixed around $500$K by latent heating, as shown in Fig.~\ref{fig:escape-example}d. This condensation temperature $T_{\mathrm{sat}}$ again can be determined by the subsaturation at the magma ocean surface (Eq.~\ref{eq:day-predict-Tcond}).
We take $T_{\mathrm{sat}}$ as an estimation of $T$ in Eq.~(\ref{eq:night-predict-massloss}). The mass flux turned upward around antistellar point can be approximated as
\begin{equation}
  \label{eq:night-predict-Phiantistellar-terminator} \frac{\Phi_{\mathrm{antistellar}}}{\Phi_{\mathrm{terminator}}}\sim \exp\left(-\frac{\pi ga}{2\sqrt{2C_p\bar{T}}\sqrt{2\pi RT_{\mathrm{sat}}}}\right).
\end{equation}
Unlike $\Phi_{\mathrm{terminator}}$, this factor has a stronger dependence on the planetary size than on temperature. The predicted $\left.\Phi_{\mathrm{antistellar}}\right/\Phi_{\mathrm{terminator}}$ is shown by red dashed curves in Fig.~\ref{fig:daynight-escape-predict}(f), in comparison with diagnosed ones (black solid). 

In the last stage, vapor transported from day-side deviates upward and escapes from the planet. In the parameter regime we explore here, internal energy alone typically is not enough for vapor to escape the gravity field ($C_pT< ag$), meaning that part of the mass flux has to condense and give their energy to the rest mass flux. As shown in section~\ref{sec:appendix-vertical-escape-night}, the moist energy $E_{\mathrm{moist}}\equiv w(r)^2/2+C_pT(r)+\Psi(r)+L\ln(\Phi(r))$ is conserved ($\Phi$ is vertical mass flux, $w$ is flow speed, and $\Psi$ is the gravity potential). Equating the $E_{\mathrm{moist}}$ at the surface $r=a$ and at infinity $r=\infty$ provides a constraint to the condensation ratio
\begin{eqnarray}
  \label{eq:night-predict-escaperatio}
  \Psi(a) +L\ln(\Phi(a))&=&\Psi(\infty)+L\ln(\Phi(\infty)).\nonumber\\
  \frac{\Phi(\infty)}{\Phi(a)}&=&\exp\left(-\frac{ga}{L}\right)
\end{eqnarray}
Here, we ignore the contribution from kinetic and internal energy, as they are small compared to the gravity potential in most cases we studied here.
Combining Eq.~(\ref{eq:night-predict-Phiterminator}), Eq.~(\ref{eq:night-predict-Phiantistellar-terminator}) with Eq.~(\ref{eq:night-predict-escaperatio}) and realizing $\Phi(a)=\Phi_{\mathrm{antistellar}}$ gives the final estimation of night-side escape flux
\begin{eqnarray}
&~&\Phi_{\mathrm{night}}=\frac{2\pi a A \sin\theta_b\sqrt{R\bar{T}}}{g}\times \exp \left(-\frac{B}{\bar{T}}-\frac{\pi ga}{4\sqrt{\pi C_pR\bar{T}T_{\mathrm{sat}}}}-\frac{ga}{L}\right).
  \label{eq:night-predict-Phi}
\end{eqnarray}
Fig.~\ref{fig:daynight-escape-predict}d shows the above estimation for various $\{T_0,M_p\}$, in comparison with the results from the full transport and escape model (Fig.~\ref{fig:daynight-escape-predict}c). Their resemblance indicates the success of Eq.~(\ref{eq:night-predict-Phi}). In fact, one more simplification can be made: the last term within the exponential ($\exp(ga/L)$) is always $O(1)$, and omitting it won't introduce significant bias (not shown).

\begin{figure*}
  \centering
\includegraphics[width=0.75\textwidth]{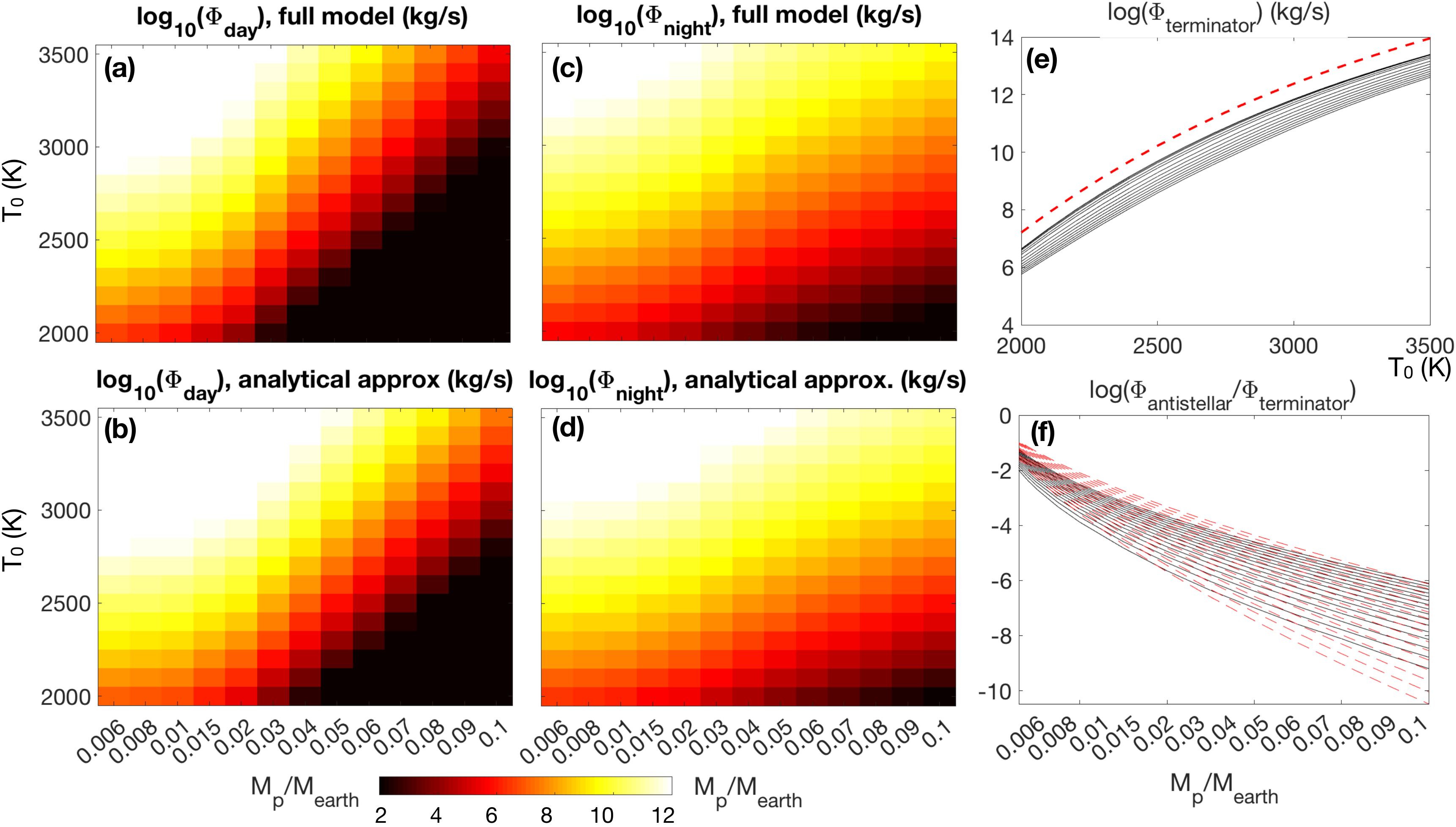}
  \caption{Same as Fig.~\ref{fig:daynight-escape-predict}, except for SiO-dominant atmosphere. Parameters used in the SiO escape calculation are summarized in Table.~\ref{tab:parameters}.}
  \label{fig:daynight-escape-predict-sio}
\end{figure*}

The same formula can bee applied to a SiO-dominant atmosphere. As shown in Fig.~\ref{fig:daynight-escape-predict-sio}, the day-side and night-side escape flux from the full escape model are reasonably predicted.

By equating the day-side escape flux $\Phi_{\mathrm{day}}$ (Eq.~\ref{eq:day-predict-Phi}) and the night-side escape flux $\Phi_{\mathrm{night}}$ (Eq.~\ref{eq:night-predict-Phi}), we can deduce the criteria dividing the regime of day-side dominant escape and the regime of night-side dominant escape.
\begin{eqnarray}
&&\frac{\Phi_{\mathrm{day}}}{\Phi_{\mathrm{night}}}=\frac{A_{\mathrm{sat}}}{4 A R^3} \frac{ g^3a^3}{ T_c^{5/2} \bar{T}^{1/2}}\frac{(1-\cos\theta_b)}{\sin\theta_b}\times\nonumber\\ &&~~\exp\left(-\frac{B_{\mathrm{sat}}}{T_c}+\frac{B}{\bar{T}}+\frac{\sqrt{\pi} ga}{4\sqrt{ C_pR\bar{T}T_{\mathrm{sat}}}}+\frac{ga}{L}\right)=1,
\end{eqnarray}
where $T_c$ decays exponentially with $ga$ as given by Eq.~(\ref{eq:day-predict-Tc}). This first term within the exponential (second line) dwarfs all others in the above equation, therefore, night-side escape dominates day-side escape on relatively large planets. Due to the low escape rate and large mass reservoir, larger planets should have a longer lifetime and thus be more easily detected.

\section{Conclusion}

In this work, we attempt to understand the first-order anisotropy of the mineral escape flow from ultra-hot rocky planets, a necessary step to explaining why some planets have trailing dusty tails, some have leading tails, and others have both \citep{Brogi-Keller-Juan-et-al-2012:evidence, Rappaport-Barclay-DeVore-et-al-2014:koi, Sanchis-Ojeda-Rappaport-Palle-et-al-2015:the, Budaj-Kocifaj-Salmeron-et-al-2015:tables}. Instead of running a multi-dimensional hydrodynamic escape model, we depict the flow field using three 1D models: one for day-side escape, one for horizontal transport from day-side to night-side and one for night-side escape. We attempt to capture, in this framework, the essential physical processes of mineral vapor escape, such as the phase changes, the exchanges in mass with the planetary surface, and the orders of magnitude pressure drop from day-side to night-side, which could be challenging for multidimensional simulations.

Regarding the first question on the explanation of trailing tails of disintegrating rocky planets, we demonstrate, using this first-order anisotropic model, that escape can occur not only on the ultra-hot day-side but also on the cold night-side. This night-side escape flow is fed by mineral vapor transport from the day-side, and can be even stronger than the day-side escape when the planet is relatively cool and large.
The night-side escape flow would naturally trail behind the planet by conserving angular momentum, and the condensation in night-side escaping flow provides an alternative explanation for the trailing tails observed by \citet{Brogi-Keller-Juan-et-al-2012:evidence, Rappaport-Barclay-DeVore-et-al-2014:koi, Sanchis-Ojeda-Rappaport-Palle-et-al-2015:the, Budaj-Kocifaj-Salmeron-et-al-2015:tables}, without considering radiation pressure proposed and discussed by \citet{Rappaport-Levine-Chiang-et-al-2012:possible, Rappaport-Barclay-DeVore-et-al-2014:koi, Sanchis-Ojeda-Rappaport-Palle-et-al-2015:the} and \citet{Lieshout-Min-Dominik-2014:dusty}. Planets dominated by the day-side escape should be hot and small, while the escape rate is high (Fig.~\ref{fig:escape-summary}a,b). A large escape rate and small planetary size would lead to an extreme short lifetime, making it unlikely to observe such planets. 

Whether radiation pressure plays an important role in forming the trailing tails depends on the dust chemical composition. Less volatile dusts made of Al$_2$O$_3$, [Fe,Mg]SiO$_3$ or [Fe,Mg]$_2$SiO$_4$ may complete multiple orbital cycles before sublimation. Under the impact of radiation pressure and gravity, dusts would travel along a trajectory that is deviated from that of the remaining gas (impacted by both pressure gradient force and gravity) and the planet (governed by gravity alone). Over time, radiation pressure would make dusts lag behind the planet in orbit \citep{Rappaport-Barclay-DeVore-et-al-2014:koi, Sanchis-Ojeda-Rappaport-Palle-et-al-2015:the, Lieshout-Min-Dominik-2014:dusty}. However, dusts made of more volatile species, such as sodium, SiO, Mg, Fe, tend to melt in a sublimation time between less than a second and a few hundred seconds \citep{Sanchis-Ojeda-Rappaport-Palle-et-al-2015:the}. That would not allow the dusts to travel far from where they form before sublimating. Instead, the constant mass exchange between condensed phase and gas phase guarantees that gas and dusts travel along the same trajectory. A mass-weighted bulk radiation pressure coefficient can then be used to describe the magnitude of radiation pressure exerts on the gas and dust as a whole. Since radiation pressure seems to be strong only for particles around 0.1-1~$\mu$m \citep{Lieshout-Min-Dominik-2014:dusty} and most of the mass stays in the gas phase, this bulk radiation pressure coefficient is likely to be small due to the dilution of the large amount gas phase materials. Therefore, radiation pressure is likely to play a less important role here, and gravity and the pressure gradient force, instead, determines trajectory of the escape flow. As far as we see, there is no clear constraint determining whether the dust would be more volatile-enriched or volatile-depleted compared to its source (i.e., a magma ocean), because the species that have high equilibrium pressure at the surface and thus can thus escape faster would also sublime more easily aloft. After all, volatile and non-volatile compositions could make comparable contributions in mass to the dust tail. This is something to be examined in future theoretical and observational works.



Useful inferences on planetary mass (the second question) could be obtained from the tail properties. Our results suggest that, although the total escape flux is affected by both temperature and planetary mass, the partition between day-side and night-side escape seems to be mainly controlled by the planetary mass. Therefore, with all other factors fixed, a larger and heavier planet is more likely to followed by a trailing tail, and vice versa, in absence of radiation pressure. Under the impact of radiation pressure, day-side escape flow could potentially be turned around and thereby trail behind the planet. A trailing tail, therefore, can be a result from radiation pressure or from the pressure gradient between day-side and night-side. This degeneracy makes it hard to constrain the masses of planets with trailing tails, unless we know about the radiation pressure coefficients \citep[this further requires knowing the composition and size of dust particles,][]{Lieshout-Min-Dominik-2014:dusty}. However, in order for a planet to have a leading tail, the day-side escape flow has to be significant, and that can put an upper limit on the planetary mass. Applying this constraint to K2-22b, a disintegrating planet with a leading tail and a 2100~K substellar point temperature \citep{Sanchis-Ojeda-Rappaport-Palle-et-al-2015:the}, suggests an upper bound of planetary mass around 0.03 earth mass (based on Fig.~\ref{fig:escape-summary} and Fig.~\ref{fig:escape-summary-sio}, the sodium and SiO calculation suggest roughly the same bound). 
After the chemical compositions of escape flows are identified through spectrometry by future missions, we may be able to better constrain the planet size from the total escape flux and the partition between leading and trailing tails. 

The advantage of this idealized framework is that it helps us to build up physical understandings of the escape process and offers guidance in deriving analytical approximations of the day-side and night-side escape rate, which can be used to explore a wide range of parameters such as chemical composition, planetary size, planetary gravity and surface temperature. For FUV-driven hydrogen escape, \citet{Watson-Donahue-Walker-1981:dynamics} derived a formula for the energy-limit escape rate. However, mineral vapor escape from rocky planets is far from this regime \citep{Perez-Becker-Chiang-2013:catastrophic}, preventing us from directly applying the energy-limit escape formula.
For escape from day-side, we provide a new method to compute the exact escape rates semi-analytically and to estimate the escape rate pure analytically, using conservation laws. This avoid doing iterations and integrations as in the original work \citep{Lehmer-Catling-Zahnle-2017:longevity}. For the night-side escape, we obtain an analytical approximation solution. These analytical approximations accurately reproduce the escape rates from the full hydrodynamic model for both sodium and SiO-dominant atmosphere (Fig.~\ref{fig:daynight-escape-predict} and Fig.~\ref{fig:daynight-escape-predict-sio}), and may be applied in studies on planetary evolution and planetary spectrometry.

\acknowledgments
We thank Prof. Andrew Ingersol, Prof. Edwin Kite, Prof. Brian Farrell and Prof. Ming Cai for insightful discussions. FD is supported by NASA grants NNX16AR86G and 80NSSC18K0829, and WK is supported by the Lorenz/Houghton Fellowship in MIT while doing this work.

\appendix

As sketched in Fig.~\ref{fig:schematics}, our model is composed of three parts: a day-side hydrodynamic escape models (section~\ref{sec:appendix-vertical-escape-day}), and a horizontal transport model calculating the day-side to night-side mass transport induced by the pressure gradient force (section~\ref{sec:appendix-horizontal-transport}), connected to a night-side hydrodynamic escape model (section~\ref{sec:appendix-vertical-escape-night}). 

Phase change is considered explicitly in all these processes. This is necessary, because, unlike hydrogen or water vapor, which remain unsaturated or saturated all the way through, the mineral gas could undergo multiple transitions between subsaturated and saturated state depending on the geometry and external energy sources. Saturated regions are demonstrated by gray shading in Fig.~\ref{fig:schematics}. The mineral gas starts its journey highly subsaturated when leaving the magma ocean, and then it can either escape directly from the day-side or follow the surface to the night-side before escaping from there. Whichever pathway it takes, the gas could become saturated through adiabatic expansion. The expansion is partially caused by the acceleration induced by the pressure gradient force (this applies to both vertical escape and horizontal transport), and partially caused by the geometric expansion along each pathway. For gas taking the day-side escape pathway, geometry always causes expansion as the escape cross section increases with the square of the distance from the planet $r$ (demonstrated by a cone drawn in dashed lines in Fig.~\ref{fig:schematics}). For gas taking the night-side escape pathway, geometry causes expansion before the terminator as the transport circumference increases along the path; beyond the terminator, the transport circumference starts to shrink, compressing the gas and turning it subsaturated.

Vertical escape of condensible matters have been considered in previous works. \citet{Lehmer-Catling-Zahnle-2017:longevity} has extended the 1D vertical escape model for saturated water vapor escape by replacing the dry air state equation with the Clausius-Clapeyron relation. However, we did not directly borrow their model, because 1) we need to explicitly consider the multiple conversions between saturated and subsaturated states of the mineral gas flow, and 2) we want to account for the condensation-induced attenuation of mass flux, which, in their calculation, is considered as conserved. \citet{Perez-Becker-Chiang-2013:catastrophic} considers the mineral gas escape from the day-side as we do here. But instead of explicitly accounting for the phase transitions, they prescribe a condensation profile. They found that the escape flux could vary by orders of magnitude when different condensation ratio profiles are prescribed (Fig.~4 in their paper), motivating us to explicitly discuss the condensation process.

In the rest of the appendix, we will present the details of the three components of our model. 


\section{1D hydrodynamic escape model for the day-side}
\label{sec:appendix-vertical-escape-day}
Since Parker's work in 1960s \citep{Parker-1965:dynamical}, 1D hydrodynamic escape model has been used to calculate the hydrogen escape from early planetary atmospheres and hot Jupiters, e.g., \citet{Kasting-Pollack-1983:loss, Watson-Donahue-Walker-1981:dynamics, Yelle-2004:aeronomy, Tian-Toon-Pavlov-et-al-2005:transonic, Murray-Clay-Chiang-Murray-2009:atmospheric, Zahnle-Catling-2017:cosmic, Zahnle-Kasting-Pollack-1990:mass} and the stream vapor escape from early icy moons and exoplanets \citet{Lehmer-Catling-Zahnle-2017:longevity, Arnscheidt-Wordsworth-Ding-2019:atmospheric}. These works are summarized by review papers, \citet{Owen-2019:atmospheric} and \citet{Lammer-Kasting-Chassefiere-et-al-2008:atmospheric}.

Our model is built upon \citet{Lehmer-Catling-Zahnle-2017:longevity}, but we will add an adiabatic escape before flow becomes saturated. The governing equations here include a mass continuity equation, a momentum equation and an energy conservation equation. The first two can be written as follows,
\begin{eqnarray}
  &~&\frac{d\Phi}{dr}\equiv \frac{d}{dr}(\rho w \Sigma) = 0\label{eq:day-vert-mass-cont}\\
  &~&w\frac{dw}{dr}=-\frac{1}{\rho}\frac{dP}{dr}+\frac{d\Psi}{dr},\label{eq:day-vert-momentum-cont}
\end{eqnarray}
where $\Phi$ refers to the mass flux, $w$ to vertical velocity, $D$ to condensation, $\rho$ to atmospheric density, $r$ to denote the distance from the center of the planet, and $g,a$ to denote the surface gravity and the radius of the planet. Parameters used in this study is summarized in Table.~\ref{tab:parameters}. $\Sigma$ is the cross section of the escaping flow. For the day-side, we assume the flow is radial, and thus $\Sigma$ increases with the square of $r$. These assumptions are supported by multi-dimensional hydrodynamic escape calculation done by \citet{Debrecht-Carroll-Nellenback-Frank-et-al-2019:photoevaporative}.
\begin{equation}
    \label{eq:S-r-day}
    \Sigma=\left.\Sigma\right|_{r=a}\left(\frac{r}{a}\right)^2
  \end{equation}
$\Psi$ denotes the gravity potential
   \begin{equation}
    \label{eq:Gravity-potential-day}
 \Psi(r)=-\frac{ga^2}{r}-\frac{3GM_*r^2}{2d^3}= -\frac{ga^2}{r}-\frac{3}{2}\Omega^2r^2
\end{equation}
The second term in the above formula is the tidal term. $M_*$ is the mass of the host star, $d$ is the distance between the star and the planet, and $\Omega$ is the orbital angular speed. This tidal term would turn the gravity force toward the host star beyond Roche lobe $r_{\mathrm{Roche}}=(M_p/3/M_*)^{1/3}d$, where $M_p=ga^2/G$ is the mass of the planet. This has been shown to be able to significantly enhance the escape flux \citep{Murray-Clay-Chiang-Murray-2009:atmospheric}.

The third equation is the equation of state. When the flow is undersaturated, the equation of state is the conservation of potential temperature $\Theta$ (or entropy equivalently),
\begin{equation}
  \label{eq:day-vert-Theta}
  \frac{d\Theta}{dr}\equiv \frac{d}{dr}(TP^{-\kappa})=0
\end{equation}
where $\kappa=R/Cp$. For sodium, $\kappa=2/5$.
When the flow becomes saturated, the equation of state is the Clausius-Clapeyron relation
\begin{equation}
  \label{eq:Psat}
  P_{\mathrm{sat}}(T_s)=A_{\mathrm{sat}}\exp\left(-\frac{B_{\mathrm{sat}}}{T_s}\right),
\end{equation}
where $A_{\mathrm{sat}}=10^{9.54}$~Pa$=10^{10.54}~Ba$, $B_{\mathrm{sat}}=12070.4$~K for sodium vapor \citep{Bowles-Rosenblum-1965:vapor}.

As Eq.~(\ref{eq:day-vert-mass-cont}) shows, we assume no mass loss even when condensation happens, as in \citet{Lehmer-Catling-Zahnle-2017:longevity}. This assumption is proper for sodium vapor escape on the day-side, because the volatile sodium droplets would absorb the stellar radiation from the day-side and quickly get re-evaporated. Actually, a sodium droplet under radiative equilibrium with the stellar radiation $S_0=\sigma T_0^4$ would have a temperature of $T_{\mathrm{droplet}}=(A_{\mathrm{droplet}}/4)^{1/4}T_0$, where $A_{\mathrm{droplet}}=0.1$ is the reflectivity of liquid sodium \citep{Barnett-Gentry-Jackson-et-al-1986:emissivity}. This would lead to an evaporation flux of $J=P/\sqrt{2\pi RT_{\mathrm{droplet}}}$ per unit surface area. A droplet of  $r_{\mathrm{droplet}}=10$~$\mu$m would completely vaporize within $\tau_{\mathrm{droplet}}=r_{\mathrm{droplet}}\rho_{\mathrm{droplet}}\sqrt{2\pi R T_{\mathrm{droplet}}}/P_{\mathrm{sat}}(T_{\mathrm{droplet}})\sim0.00078$~s, with $T_0=2500$~K and $\rho_{\mathrm{droplet}}=968$~kg/m$^3$. That means the droplet re-evaporates almost immediately after it reaches its equilibrium temperature. Given the high volatility of sodium, a better estimation of sodium dust's lifetime under the sun is given by energy budget $\tau_{\mathrm{droplet}}=\left.4L\rho_{\mathrm{droplet}}r_{\mathrm{droplet}} \right/(AS_0)$, which gives 0.75~s. Such a short sodium droplet lifetime indicates that the mass flux of vapor is indeed almost conserved. This energy source is taken into account by ignoring the mass loss in the mass continuity equation. However, as to be discussed in section~\ref{sec:appendix-vertical-escape-night}, the re-evaporation of night-side escape is negligible in absence of stellar radiation. 

The property of escape flow before condensation can be solved from Eq.~(\ref{eq:day-vert-mass-cont}), Eq.~(\ref{eq:day-vert-momentum-cont}) with Eq.~(\ref{eq:day-vert-Theta}) and Eq.~(\ref{eq:Psat}), given the initial condition at the surface,
\begin{eqnarray} 
  T|_{r=a}&=&T_s \label{eq:boundary-T-day}\\
  P|_{r=a}&=&P_{\mathrm{chem}}(T|_{r=a}). \label{eq:boundary-P-day}
\end{eqnarray}
Here, we assume the surface vapor temperature $T|_{r=a}$ is the same as the magma ocean and the surface vapor pressure $P|_{r=a}$ is in chemical equilibrium with the ocean. In reality, the vapor pressure would be lower because vaporization requires the atmosphere to be subsaturated at the surface.
We take the form of $P_{\mathrm{chem}}$ from \citet{Castan-Menou-2011:atmospheres}
\begin{equation}
  P_{\mathrm{chem}}(T_s)=A\exp(-\frac{B}{T_s}) \cdot\mathbbm{1}[T_s>T_m],\label{eq:Peq}
\end{equation}
where $A_{\mathrm{chem}}=10^{9.6}$~Pa$=10^{10.6}$~Ba, $B_{\mathrm{chem}}=38000$~K. This equilibrium pressure is much lower than the saturated pressure for the same temperature, because sodium only takes a small fraction in the magma ocean, and the equilibrium pressure drops following Henry's law. $\mathbbm{1}[T_s>T_m]=1$ within the magma ocean when the surface temperature $T_s$ is higher than the melting temperature of the magma $T_m=1673$ \citep{Kite-Fegley-Schaefer-et-al-2016:atmosphere} and $\mathbbm{1}[T_s>T_m]=0$ elsewhere.

As pointed out by \citet{Kasting-Pollack-1983:loss, Watson-Donahue-Walker-1981:dynamics, Yelle-2004:aeronomy, Tian-Toon-Pavlov-et-al-2005:transonic, Murray-Clay-Chiang-Murray-2009:atmospheric, Zahnle-Catling-2017:cosmic, Zahnle-Kasting-Pollack-1990:mass}, the above equation is singular around transonic point, and requiring a smooth transition across the singular point puts constraints on the flow properties there. Because of the unforeseeable condensation, our situation is slightly more complicated than the dry adiabatic escape \citep{Pierrehumbert-2010:principles} and the pure steam escape \citep{Lehmer-Catling-Zahnle-2017:longevity}. Therefore, we need to consider two possibilities: flow turns supersonic while it is undersaturated, versus flow turns supersonic while it is saturated.

For the first case, some manipulations of Eq.~(\ref{eq:day-vert-mass-cont}), Eq.~(\ref{eq:day-vert-momentum-cont}) and Eq.~(\ref{eq:day-vert-Theta}) lead to
\begin{equation}
  \label{eq:transonic-eq-dry}
  \left(w^2-\frac{RT}{1-\kappa}\right)\frac{d\ln w}{dr}=-\frac{d\Psi}{dr}-\frac{RT}{\kappa-1}\frac{d\ln \Sigma}{dr}.
\end{equation}
To avoid jumps in $w$, the left hand side and right hand side have to vanish simultaneously at the transonic point, which leads to the transonic condition below.
\begin{equation}
  w_c^2=\frac{RT_c}{1-\kappa}=\left.\frac{d\Psi}{d\ln \Sigma}\right|_c\label{eq:transonic-cond-dry}
\end{equation}
Subscript $(\cdot)_c$ indicates that certain variables are evaluated at the transonic point.

For the second case (i.e., flow becomes saturated before transonic point), Eq.~(\ref{eq:day-vert-mass-cont}), Eq.~(\ref{eq:day-vert-momentum-cont}) and Eq.~(\ref{eq:Peq}) give
\begin{equation}
  \label{eq:transonic-eq-moist}
   \left(w^2-\frac{RTB_{\mathrm{sat}}}{B_{\mathrm{sat}}-T}\right)\frac{d\ln w}{dr}=-\frac{d\Psi}{dr}-\frac{RTB_{\mathrm{sat}}}{B_{\mathrm{sat}}-T}\frac{d\ln \Sigma}{dr},
 \end{equation}
 which, in turn, leads to a different transonic condition
 \begin{equation}
   \label{eq:transonic-cond-moist}
   w_c^2=\frac{RT_cB_{\mathrm{sat}}}{B_{\mathrm{sat}}-T_c}=\left.\frac{d\Psi}{d\ln \Sigma}\right|_c
 \end{equation}
 
 With the transonic conditions, one may make a guess of $w_c$, solve for $T_c,\ r_c$, and then integrate the momentum equation (Eq.~\ref{eq:day-vert-momentum-cont}) downward to surface to match the bottom boundary condition given by Eq.~\ref{eq:boundary-T-day} and Eq.~\ref{eq:boundary-P-day}, as done in \citet{Lehmer-Catling-Zahnle-2017:longevity}. We here do it in a slightly different way. Instead of making guesses and doing iterations, we take advantage of conservation laws to directly solve for the flow properties at the transonic point and thereby the escape flux, saving us from going through the iteration.

 We first assume that escape flow is still undersaturated when turning supersonic. For an undersaturated flow, there are three conserved quantities, potential temperature $\Theta$, mass flux $\Phi$ and energy density
 \begin{equation}
   \label{eq:vert-day-dry-energy}
   E=\frac{1}{2}w^2+C_pT+\Psi.
 \end{equation}
 The conservation of $\Theta$, $\Phi$ and $E$ before the transonic point yields
 \begin{eqnarray}
   T_cP_c^{-\kappa}&=& T_sP_s^{-\kappa}\label{eq:vert-day-conserve-dry-Theta}\\
   \frac{P_c}{T_c}w_c\Sigma(r_c)&=&\frac{P_s}{T_s}w_s\Sigma(a)\label{eq:vert-day-conserve-Phi}\\
   \frac{1}{2}w_c^2+C_pT_c+\Psi(r_c)&=&\frac{1}{2}w_s^2+C_pT_s+\Psi(a), \label{eq:vert-day-conserve-dry-E}
 \end{eqnarray}
 where subscript $(\cdot)_s$ denotes quantities evaluated at the surface, $r=a$. Combining Eq.~(\ref{eq:vert-day-conserve-dry-Theta}-\ref{eq:vert-day-conserve-dry-E}) with the transonic condition (Eq.~\ref{eq:transonic-cond-dry}), we get five equations, from which the five unknowns, $r_c,\ w_c,\ T_c,\ P_c$ and $w_s$, can be solved.

 We then examine whether the escape flow is indeed undersaturated all the way to the supersonic point by comparing $P_c$ with the saturated vapor pressure $P_{\mathrm{sat}}(T_c)$. If the escape flow turns out to be saturated before the transonic point, we can only apply the $\{\Theta, \Phi, E\}$ conservation laws before saturation occurs. 
  \begin{eqnarray}
   T_sP_s^{-\kappa}&=& T_{\mathrm{sat}}P_{\mathrm{sat}}^{-\kappa}\label{eq:vert-day-conserve-moist-Theta}\\
   \frac{P_s}{T_s}w_s\Sigma(r_s)&=&\frac{P_{\mathrm{sat}}}{T_{\mathrm{sat}}}w_{\mathrm{sat}}\Sigma(r_{\mathrm{sat}})\label{eq:vert-day-conserve-moist-Phi}\\
   \frac{1}{2}w_s^2+C_pT_s+\Psi(r_s)&=&\frac{1}{2}w_{\mathrm{sat}}^2+C_pT_{\mathrm{sat}}+\Psi(r_{\mathrm{sat}}), \label{eq:vert-day-conserve-moist-E}
  \end{eqnarray}
  where $(\cdot)_{\mathrm{sat}}$ denotes quantities evaluated at the saturation level $r_{\mathrm{sat}}$. Beyond $r_{\mathrm{sat}}$, $E$ and $\Theta$ are no longer conserved, instead we have the Clausius-Clapeyron relation (Eq.~\ref{eq:Psat}) and a Bernoulli function type of conserved quantity (can be derived from Eq.~\ref{eq:day-vert-momentum-cont} by replacing $\rho,\ P$ with $T$ using Clausius-Clapeyron relation and ideal gas law) linking $r_{\mathrm{sat}}$ and $r_c$
  \begin{equation}
    \label{eq:vert-day-conserve-moist-B}
    \Psi(r_c)+\frac{1}{2}w_c^2+RB_{\mathrm{sat}}\ln T_c=\Psi(r_{\mathrm{sat}})+\frac{1}{2}w_{\mathrm{sat}}^2+RB_{\mathrm{sat}}\ln T_{\mathrm{sat}}.
  \end{equation}
  Then, with eight equations, namely, Eq.~(\ref{eq:transonic-cond-moist}), Eq.~(\ref{eq:vert-day-conserve-Phi}), Eq.~(\ref{eq:Psat}) and Eq.~(\ref{eq:vert-day-conserve-moist-Theta}-\ref{eq:vert-day-conserve-moist-B}), eight unknowns, $r_c$, $T_c$, $P_c$, $w_c$, $w_s$, $r_{\mathrm{sat}}$, $P_{\mathrm{sat}}$, $T_{\mathrm{sat}}$, can be solved.
  
  With the flow properties at the transonic point, we can directly calculate the escape flux
  \begin{equation}
    \label{eq:vert-day-mass-flux}
    M=\frac{P_c}{RT_c}w_c\Sigma(r_c),
  \end{equation}
  as well as other conserved quantities. Then, flow properties at any give location $r$ can be solved from the conservation laws, which are all algebraic equations.

  Since the surface temperature decreases away from the substellar point, and the dependence of $P$ on $T$ is highly nonlinear (exponential), we divide the region covered by magma ocean into $N_\theta$ isothermal sectors and calculate the total escape rate by summing over the escape from each sector. Direct escape beyond magma ocean is ignored. In most cases, we found $N_\theta=5$ is good enough. Too few sectors will lead to underestimation of the escape flux.


Absorption of UV flux from the star trigger ionization in the interior of the escape flow, which in turn leads to heating and stronger escape. Simulating this would require to explicitly couple ionization, radiative transfer models with the hydrodynamic escape model as done in \citet{Murray-Clay-Chiang-Murray-2009:atmospheric}, and is out of our scope here. Actually, as pointed out by \citet{Perez-Becker-Chiang-2013:catastrophic}, energy deposition by photoionization is not necessary for driving an escape flow, because most of the disintegrating terrestrial planets are small in size and hence have weak gravity. To get an order of magnitude estimate, we take the energy limit escape given by \citet{Watson-Donahue-Walker-1981:dynamics},
\begin{equation}
  \label{eq:energy-limit-escape}
  \Phi_{\mathrm{energy-lim}}=\frac{\eta F^\downarrow_{\mathrm{uv}}(\infty) \Sigma_1}{ga}.
\end{equation}
In the above equation, $\eta=\frac{h\nu_0-E_{\mathrm{ionize}}}{h\nu_0}$ is the efficiency for FUV flux with mean photon energy of $h\nu_0=20$~eV to be converted to thermal energy in the vapor. Ionization takes energy of $E_{\mathrm{ionize}}$, this part of energy cannot be thermalized until recombination (ignored) happens. $\Sigma_1$ is the cross section of the escape flow where the FUV flux is absorbed (optical depth equals one). Since we only attempt to get an estimation for orders of magnitude, we let $\Sigma_1=\pi a^2$ for simplicity.
The downward FUV flux $F^\downarrow_{\mathrm{uv}}(\infty)$ at the top of the atmosphere is set to $0.45$ W/m$^2$. This value has been used to calculate the hydrogen escape from hot jupiter whose host star is not a T-tauri star \citep{Murray-Clay-Chiang-Murray-2009:atmospheric}.
The UV-induced escape turns out to be around $10^6$~kg/s, negligible compared to the thermally driven escape for most cases except those marked by the dark red color in Fig.~\ref{fig:escape-summary}a and Fig.~\ref{fig:escape-summary-sio}a.

\section{1D horizontal transport model.}
\label{sec:appendix-horizontal-transport}

Our horizontal transport model is built upon \citet{Ingersoll-Summers-Schlipf-1985:supersonic}, which was used to calculate how SO$_2$ is transported from the day-side to night-side of Io. We will not repeat the derivations; rather, we will present the equations and highlight the changes we make. Interested readers are referred to \citet{Ingersoll-Summers-Schlipf-1985:supersonic} for more details. The governing equations include the conservation of vertically integrated mass flux, momentum flux and energy flux.
\begin{eqnarray}
  &~&  \frac{1}{a \sin\theta}\frac{d}{d\theta}(VP/g\sin\theta) = F-D \label{eq:horiz-mass-cont}\\
  &~&  \frac{1}{a \sin\theta}\frac{d}{d\theta}((V^2+\beta C_pT)P/g\sin\theta) = \beta C_p TP\cot\theta/a +(\min\{F,0\}-D)V \label{eq:horiz-momentum}\\
  &~&  \frac{1}{a \sin\theta}\frac{d}{d\theta}((V^2/2+ C_pT) V P/g\sin\theta) =  DL + (\min\{F,0\}-D)(V^2/2+C_pT) + \max\{F,0\} C_pT_s(\theta). \label{eq:horiz-energy}
\end{eqnarray}
We use $V$ for the flow speed, $P$ for surface atmosphere pressure, $T$ for average air temperature, $T_s$ for surface temperature, $F$ for the mass exchange with the surface (positive for vaporization), $L$ for latent heat, $a$ for the planet radius and $g$ for the gravity acceleration rate of the planet, respectively. Parameters used in this study is summarized in Table~\ref{tab:parameters}. The mass exchange $F$ is positive (negative) when the atmosphere pressure $P$ is lower (higher) than $P_{\mathrm{chem}}$, the chemical equilibrium pressure with the magma ocean.
\begin{eqnarray}
   \label{eq:F-P}
  F&=&\begin{cases}\frac{\alpha P}{\sqrt{2\pi RT}}\left(\frac{P_{\mathrm{chem}}(T_s)}{P}-1\right) & \mathrm{within~magma~ocean}\\
    \min\left\{\frac{\alpha P}{\sqrt{2\pi RT}}\left(\frac{P_{\mathrm{sat}}(T_s)}{P}-1\right),D \right\} & \mathrm{out~of~magma~ocean}\end{cases}
\end{eqnarray}
Physically, the above formula means that a proportion $\alpha$ of the total number of collisions per unit time per area $P/\sqrt{2\pi RT}$ act to restore the vapor pressure toward the equilibrium pressure $P_{\mathrm{chem}}(T_s)$. With $\alpha=1$ (default), every collision counts. We ignore the fact that the exchange efficiency is usually less than one; this becomes increasingly important when the surface temperature is relatively warm \citep{Haynes-Tro-George-2002:condensation}. Meanwhile, we also ignore the turbulence induced exchange in the surface boundary layer, which was found to be unimportant in \citet{Ingersoll-1989:io} in the context of Io and \citet{Castan-Menou-2011:atmospheres} in the context of tidally-locked super earth. With the turbulence drag coefficient $C_d$ set to $0.01$ as in \citet{Ingersoll-Summers-Schlipf-1985:supersonic}, the characteristic speed corresponding to the turbulent exchange is only $C_dV\lesssim 20$~m/s. This is much smaller than that corresponding to $F$, which is close to the sound speed. 

Within magma ocean, the equilibrium pressure is determined by the chemical equilibrium between the gas phase and the dissolved phase (Eq.~\ref{eq:Peq}). 
Out of magma ocean, the equilibrium pressure instead is given by the Clausius-Clapeyron relation (Eq.~\ref{eq:Psat}).
Another change is that there is no sodium reservoir at the surface out of magma ocean, therefore, the outward mass flux $F$ cannot exceed the amount of local condensation $D$. This corresponds to the $\min$ operator in Eq.~(\ref{eq:F-P}).

We assume the surface temperature is in radiative equilibrium close to the substellar point and it gradually cools down until it reaches an arbitrary minimum night-side temperature $T_{N}=50$~K\footnote{The results won't be affected by the choice of $T_N$, as long as $T_N$ is lower than the air temperature, which is above 500~K.} \citep{Castan-Menou-2011:atmospheres, Kite-Fegley-Schaefer-et-al-2016:atmosphere}.
\begin{equation}
  \label{eq:Ts}
  T_s=T_N+(T_0-T_N)\max\{\cos\theta,0\}^{1/4},
\end{equation}
$\theta$ is the angle distance from the substellar point, and $T_0$ is the substellar point. In reality, the surface temperature would deviate from the radiative equilibrium because of the greenhouse effect and blocking of solar radiation by dusts in the escape flow \citep{Perez-Becker-Chiang-2013:catastrophic}.

A major change we make here is to explicitly consider the condensation due to over-saturation. The condensation $D$ is positive when the gas pressure $P$ is higher than the saturated vapor pressure $P_{\mathrm{sat}}(T)$.
$D$ not only enters the mass equation (Eq.~\ref{eq:horiz-mass-cont}) but also the momentum and energy equations (Eq.~\ref{eq:horiz-momentum}-\ref{eq:horiz-energy}) because the condensed mass will take its momentum and energy with it so that the flow speed and the flow energy density will not be affected. Meanwhile, condensation will cause latent heating release $DL$ in the interior of the flow (Eq.~\ref{eq:horiz-energy}), preventing further temperature drop in the flow.



We integrate Eq.~(\ref{eq:horiz-mass-cont}-\ref{eq:horiz-energy}) from the substellar point to the night-side using finite difference method. When the atmosphere is undersaturated, $D$ equals zero. We can integrate forward the mass flux $\Phi=VPa\sin\theta/g$ and the dry energy density of the flow $E=V^2/2+C_pT$ knowing the surface exchange flux $F$, which in turn can be calculated from Eq.~\ref{eq:F-P} given $T_s,\ T$ and $P$ at the current grid point.
\begin{eqnarray}
  \Phi_{(n+1)}-\Phi_{(n)}&=&(a^2F_{(n)})(\cos\theta_{(n)}-\cos\theta_{(n+1)})\label{eq:horiz-unsat-Phi}\\
  E_{(n+1)}-E_{(n)}&=&(a^2F_{(n)}/\Phi_{(n)})(C_pT_{s(n)}-C_pT_{(n)}-V_{(n)}^2/2)(\cos\theta_{(n)}-\cos\theta_{(n+1)}) \label{eq:horiz-unsat-E}
\end{eqnarray}
Subscripts in the above equation denote the index of grid point.
With $\Phi_{(n+1)}$ and $E_{(n+1)}$ solved from the above equations, the number of unknowns reduces to one. The new grid point's $T_{(n+1)}$ and $P_{(n+1)}$ are linked with $V_{(n+1)}$ through
\begin{eqnarray}
  T_{(n+1)}&=&(E_{(n+1)}-V_{(n+1)}^2/2)/C_p\label{eq:horiz-unsat-T}\\
  P_{(n+1)}&=&\frac{\Phi_{(n+1)}g}{a\sin\theta_{(n+1)}V_{(n+1)}}. \label{eq:horiz-unsat-P}
\end{eqnarray}
Substituting the above equations to the finite-differenced momentum equation (Eq.~\ref{eq:horiz-momentum}) lead to 
\begin{equation}
  \label{eq:horiz-unsat-V}
  \beta C_p(T_{(n+1)}P_{(n+1)}-T_{(n)}P_{(n)})+V_{(n)}P_{(n)}(V_{(n+1)}-V_{(n)})=-\max\{F,0\}V_ng(\theta_{(n+1)}-\theta_{(n)}),
\end{equation}
from which $V_{(n+1)}$ can be solved. 

When the flow becomes saturated, $D$ is no longer zero and is unknown, but Clausius-Clapeyron relation will link $P$ with $T$, keeping the total number of the unknowns at three. To make the integration faster, we cancel $D$ in Eq.~(\ref{eq:horiz-mass-cont}-\ref{eq:horiz-energy}) to get
\begin{eqnarray}
  &~&\frac{d}{d\theta}\left[V^2/2+C_pT+ L\ln(VP\sin\theta)\right]=\frac{g}{VP}\left[LF+\max\{F,0\}(C_pT_s-C_pT-V^2/2)\right]\label{eq:horizontal-sat-1}\\
  &~&\beta C_p\frac{d}{d\theta}(TP)+VP\frac{d}{d\theta}V=-\max\{F,0\}Vg. \label{eq:horizontal-sat-2}
\end{eqnarray}
Rewriting the above equations using finite difference and combining the definition of $F$ in Eq.~\eqref{eq:F-P} and the Clausius-Clapeyron relation in Eq.~\eqref{eq:Psat} yield an algebatic equation set, from which $V,\ T$ in the next grid point can be solved knowing the current state of the flow.

The reservoir constraint in the formula for $F$ (Eq.~\ref{eq:F-P}) needs some additional care, because the upper bound for $F$ is set by the unknown condensation rate $D$. For regions within the magma ocean, the reservoir constraint is not relevant. The night-side ($\theta>90^\circ$) is also one of the unconstrained regions because the surface temperature $T_N=50K$ is cold enough to prevent any reevaporation. However, most of day-side ``land'' surface can be much warmer than the vapor on top of it. Thus, whatever condensed out of the atmosphere would be reevaporated from the surface at a rate prescribed by the saturated vapor pressure at the surface, but this reevaporation should not exceed the condensation, as the solidified surface does not continuously supply sodium.

For these regions, we first solve Eq.~\eqref{eq:horizontal-sat-1} and Eq.~\eqref{eq:horizontal-sat-2} assuming the reservoir limit has been reached, i.e., $F=D>0$. Under this limit, the mass flux $\Phi$ won't change
\begin{equation}
  \label{eq:horiz-satFD-Phi}
  \Phi_{(n+1)}=V_{(n+1)}P_{(n+1)}a\sin\theta_{(n+1)}/g=\Phi_{(n)}.
\end{equation}
 With the above mass continuity equation and the Clausius-Clapeyron relation (Eq.~\ref{eq:Psat}), unknowns in Eq.~\eqref{eq:horizontal-sat-1} and Eq.~\eqref{eq:horizontal-sat-2} reduces to $T$ and $D$, and everything about the flow at the new grid point can be solved.
Then we compare the above $F$ solution with the surface flux intended by the pressure imbalance ignoring the reservoir constraint
\begin{equation}
  \label{eq:F-intend}
F'=\frac{\alpha P}{\sqrt{2\pi RT}}\left(\frac{P_{\mathrm{sat}}(T_s)}{P}-1\right).
\end{equation}
If $F'\geq F$, then the $D=F$ assumption is valid; otherwise, we recalculate $D,\ T$ without the reservoir constraint. Solving the reservoir constrained situation first turns out to be critical for the stability of the integration, because without reservoir constraint, the intended surface flux could be gigantic and that could lead to nonphysical solutions.

As mentioned by \citet{Ingersoll-Summers-Schlipf-1985:supersonic} and \citet{Castan-Menou-2011:atmospheres}, one need to make a guess of the pressure at the substellar point $P_0$ to start the integration. A too large $P_0$ (too close to the $P_{\mathrm{chem}}$ there) will make the mass gain from the magma ocean $F$ not large enough to support the flow. In the midst of the integration, the flow will slow down and return, inconsistent with the boundary condition $V=0$ at antistellar point. A too small $P_0$ will lead to too strong sublimation and too fast acceleration so that solution will no longer exist beyond certain point. We identify these two types of mistakes and avoid them by adjusting the initial guess of $P_0$, until a smooth solution is found for the whole domain. 

No mass can pass the antistellar point by definition. We stop the integration when the speed drops below one half of the peak speed, and we expect the flow to turn upward beyond this point as sketched in Fig.~\ref{fig:schematics}.

\section{1D hydrodynamic escape model for the night-side}
\label{sec:appendix-vertical-escape-night}

The governing equations here again include the conservation of mass flux, momentum flux and energy flux.
\begin{eqnarray}
  &~&\frac{d}{dr}\Phi = -D\label{eq:night-vert-mass-cont}\\
  &~&\frac{d}{dr}(\Phi w) = -\rho \Sigma\frac{d\Psi}{dr} - \Sigma \frac{dP}{dr} -D w\label{eq:night-vert-momentum-cont}\\
  &~&\frac{d}{dr}(\Phi (w^2/2+C_pT+\Psi)) = D(L-w^2/2-C_pT+\Psi) \label{eq:night-vert-energy-cont}
\end{eqnarray}
where $D$ represents the mass loss due to condensation, and the definition of other symbols follow that in the day-side hydrodynamic model.
\begin{equation}
  \label{eq:massflux}
  \Phi=\rho w \Sigma(r)
\end{equation}
is the upward mass flux and
\begin{equation}
  \label{eq:Gravity-potential-night}
  \Psi(r)=-ga^2/r
\end{equation}
is the gravity potential.
For any given $r$, the flow cross section is given by 
\begin{equation}
  \label{eq:S-r-night}
  \Sigma= \left.\Sigma\right|_{r=a}\left(\frac{r}{a}\right)^\epsilon
  \end{equation}
where $\epsilon$ is a small real number. Other parameters used in this study is summarized in Table~\ref{tab:parameters}. As sketched in Fig.~\ref{fig:schematics}, for the night-side, we let $\epsilon=0.2$, and this corresponds to a slowly expanding night-side cone away from the planet. The night-side escape flux is relatively insensitive to $\epsilon$, unless $\epsilon$ is smaller than $0.02$. The insensitivity of escape rate to $\epsilon$ is because we fix the mass flux at the surface. As demonstrated in section~\ref{sec:analytical-night}, the ratio between the mass flux that eventually escape from the night-side and the flux we imposed at the bottom boundary of the escape model is determined by how much percentage of the escape gas needs to give their latent heating up to the remaining gas (by condensing) so that the remaining gas have enough energy to escape from the planetary gravity field. However, we do notice that, with an extremely small $\epsilon$, the escape flow would be geometrically confined to a subsonic breeze, just like what happens when a stellar pressure is imposed onto the day-side escape flow in \citet{Murray-Clay-Chiang-Murray-2009:atmospheric}.

Since the escape flow on the night-side does not receive direct radiation from the star, mineral vapor, once condensed, can hardly re-evaporate. Therefore, we think it is necessary to consider not only the latent heating release but also the mass loss and pressure drop due to condensation $D$, which is neglected in previous works, e.g., \citet{Perez-Becker-Chiang-2013:catastrophic, Lehmer-Catling-Zahnle-2017:longevity, Zahnle-Catling-2017:cosmic}. In our calculation, we will ignore the momentum exchange between dusts and the rest vapor, and thus we would expect a different speed distribution and thus a different trajectory for the dusty and gaseous tails. As shown below, the night-side mass flux at the surface is set to be equal to the mass flux of the horizontal transport flow. Ignoring mass loss due to condensation as in previous studies is equivalent to force a 100\% escape of whatever transported close to the antistellar point. This could be unphysical because we would expect zero escape if the planetary gravity is indefinitely strong.

This condensation-induced mass loss is represented by $D$ in Eq.~(\ref{eq:night-vert-mass-cont}-\ref{eq:night-vert-energy-cont}).
We let $D>0$ whenever atmospheric pressure $P$ is higher than the saturated vapor pressure $P_{\mathrm{sat}}$ (Eq.~\ref{eq:Psat}); and reevaporation ($D<0$) happens when $P<P_{\mathrm{sat}}$\footnote{Aerosols may travel at different speeds than the vapor flow, so reevaportation may cause change to the mean upward momentum of the flow. That is ignored here.}. We stop the reevaporation by turning the flow undersaturated, when the local upward mass flux $\Phi$ equals/exceeds that the flow begins with.

The initial state of the night-side escape model is connected with the final state of the horizontal transport model through mass flux and energy flux continuity. As mentioned in section~\ref{sec:appendix-horizontal-transport}, we stop the horizontal transport when the flow speed drops below $50\%$ of its peak value and force the flow turn upward. At this turning point, complex physics processes such as turbulent mixing, conversions between kinetic energy and thermal energy and phase change may occur, and fully resolving these processes would require a 3D hydrodynamic simulation that can deal with orders of magnitude pressure change in a short distance, which is beyond the scope of our work. For simplicity, we take the final energy flux and mass flux from the horizontal transport model and use it as the initial condition for the vertical escape model.
\begin{eqnarray}
  \left.VP/g(2\pi a\sin\theta)\right|_{\text{horiz},V=0.5V_{\max}}&=&\left.\Sigma (P/RT)w\right|_{\text{vert},r=a}  \label{eq:connect-mass}\\
  \left.C_pT+V^2/2\right|_{\text{horiz},V=0.5V_{\max}}&=& \left.C_pT+w^2/2\right|_{\text{vert},r=a}\label{eq:connect-energy}\\
  \left.\pi (a\sin\theta)^2\right|_{\text{horiz},V=0.5V_{\max}}&=&\left.\Sigma \right|_{\text{vert},r=a}  \label{eq:connect-area}.
\end{eqnarray}
We call the above equations the connection conditions, connecting the ``last moment'' in horizontal transport model and the ``first step'' in the vertical escape model.
This way, we conserve the energy and mass input from the horizontal flow, while allowing the high pressure formed at the anti-stellar point to slow down the horizontal flow and convert the associated kinetic energy to thermal energy.

This connection condition could lead to an initial state that is supersaturated. It is particularly likely to happen when the planetary gravity is strong, as escape flow would get trapped and become dense. If the connection condition yields an initial state $T,\ P$, where $P>P_{\mathrm{sat}}(T)$, we first let the vapor at the surface first undergoes an initial condensation without changing its $w$, so that the adjusted $T',\ P'$ follow Clausius-Clapeyron. 
\begin{equation}
  \label{eq:supersaturate-surface}
  C_pT+L\ln(P/RT)=C_pT'+L\ln(P_{\mathrm{sat}}(T')/RT').
\end{equation}
We solve the adjusted $T'$ from the above equation, and then update pressure using $P'=P_{\mathrm{sat}}(T')$.

Starting from the surface, flow is usually undersaturated, except some rare cases, where vapor starts super dense on the night-side (to be discussed later). For undersaturated escape flow, three conserved quantities can be found after some term rearrangements of Eq.~(\ref{eq:night-vert-mass-cont}-\ref{eq:night-vert-energy-cont}): mass flux $\Phi$, energy $E$ and potential temperature $\Theta$
\begin{eqnarray}
  \Phi&=&\rho w \Sigma \label{eq:dry-conserve-mass}\\
  E&=&w^2/2+C_pT+\Psi(r) \label{eq:dry-conserve-energy}\\
  \Theta&=& TP^{-\kappa}.\label{eq:dry-conserve-pt}
\end{eqnarray}
$\Phi,\ E$ and $\Theta$ can be calculated given the initial condition. 

Then $w,\ T,\ P$ can be solved from the above algebratic equations, until the flow cools down and turns saturated. The location where flow turns saturated $r_{\mathrm{sat}}$ can be calculated by solving Eq.~(\ref{eq:dry-conserve-mass}-\ref{eq:dry-conserve-pt}) together with the Clausius-Clapeyron relation (Eq.~\ref{eq:Psat}).
 
After the flow turns saturated, the atmosphere pressure $P$ is linked with temperature $T$ through Clausius-Clapeyron relation (Eq.~\ref{eq:Psat}). Besides this relation, another two conserved quantities can be derived from Eq.~(\ref{eq:night-vert-mass-cont}-\ref{eq:night-vert-energy-cont}).
\begin{eqnarray}
  E_{\mathrm{moist}}&=&w^2/2+C_pT+\Psi(r)+L\ln(\rho w \Sigma)\label{eq:moist-conserve-energy}\\
  B_{\mathrm{moist}}&=&w^2/2+\Psi(r)+RB_{\mathrm{sat}}\ln T.\label{eq:moist-conserve-bernoulli}
\end{eqnarray}
The moist energy $E_{\mathrm{moist}}$ is conserved for undersaturated flow too. $E_{\mathrm{moist}}$ and $B_{\mathrm{moist}}$ can be calculated before doing the integration, knowing $r_{\mathrm{sat}}$ and the vapor properties there. Then we can solve $w,\ P,\ T$ profiles from Eq.~(\ref{eq:moist-conserve-energy}), Eq.~(\ref{eq:moist-conserve-bernoulli}) and Eq.~\ref{eq:Psat}.

As $\Sigma$ increases with $r$, vapor becomes thinner and colder. Eventually, vapor would either condense into dusts, or it would stop behaving like fluid as the collision between molecules becomes less and less frequent. The latter becomes relevant beyond the exobase, whose definition is
\begin{equation}
  \label{eq:exobase}
  \int_{r_{exo}}^\infty n(r')\sigma_{\mathrm{collision}}dr'=1,
\end{equation}
where $\sigma_{\mathrm{collision}}$ is the collision cross section of sodium.
For simplicity, we treat both cases as condensation. That means all vapor will condense by the end. Once condensation happens, we assume the dust's kinetic energy is conserved. Those with kinetic energy higher than the highest potential along its path will be able to escape, and the rest will fall back to surface. Since the escape flow's kinetic energy always increases with $r$, an equivalent way to state the escape criteria is that dusts can escape the planetary gravity if and only if they condense beyond a $r_{\mathrm{escape}}$
\begin{equation}
  \label{eq:r-escape}
  w(r_{\mathrm{escape}})^2/2+\Psi(r_{\mathrm{escape}})=0. 
\end{equation}
With $r_{\mathrm{escape}}$, the escape flux can be determined as
\begin{equation}
  \label{eq:Phi-escape}
  \Phi_{\mathrm{escape}}=\Phi(r_{\mathrm{escape}})
\end{equation}

Similar to the horizontal transport model, singularity forms when flow turns supersonic. This has been noted by \citet{Kasting-Pollack-1983:loss, Watson-Donahue-Walker-1981:dynamics, Yelle-2004:aeronomy, Tian-Toon-Pavlov-et-al-2005:transonic, Murray-Clay-Chiang-Murray-2009:atmospheric}. Actually, the connection conditions (Eq.~\ref{eq:connect-mass}-\ref{eq:connect-area}) do not fully constrain the flow properties at the surface. Instead, there is one degree of freedom left. We thus need to make a guess for the initial speed $w|_{r=a}$. A too large initial $w|_{r=a}$ would lead to an unphysical large pressure drop due to the adiabatic cooing and/or condensation, which in turn, accelerates the flow even more. Beyond a certain point, solution will not exist any more. On the other hand, starting from too small initial $w|_{r=a}$, the flow would end up slowing down and turning backward. Only a proper initial guess allows the solution to pass the singularity smoothly. We find out this proper initial condition through a binary search.

\section{Other remarks for the model}
\label{sec:appendix-other-remarks}
We here focus on sodium dominant atmosphere, but our method can easily be applied to other less volatile components, such as SiO, Mg and even SiO$_2$, Fe, Fe$_2$SO$_4$, MgSO$_3$, Al$_2$O$_3$ \citep{Lieshout-Min-Dominik-2014:dusty}.
The orbital parameters are taken from from KIC-1255b \citep{Brogi-Keller-Juan-et-al-2012:evidence, Lieshout-Rappaport-2018:disintegrating}, as summarized in Table.~\ref{tab:parameters}.

\begin{table}[hptb!]
  \centering
  \begin{tabular}{lll}
    \hline
    Symbol & Name & Definition/Value\\
    \hline
    $M_{*}$ & mass of host star & 0.67 M$_{\mathrm{sun}}$\\
    $d$ & semi-major axis & 0.013~AU\\
    $F^\downarrow_{\mathrm{uv}}(\infty)$ & FUV flux received by the planet & 0.45~W/m$^2$ (450~erg/s/cm$^2$)\\
    $h\nu_0$ & mean energy of FUV photon & 0.20~eV\\
    $T_N$ & night-side surface temperature & 50K\\
    $T_m$ & melting temperature of magma & 1673K\\
    $\alpha$ & exchange efficiency at vapor-condensed interface & 1\\
    \hline
    \multicolumn{3}{l}{sodium-dominant atmosphere}\\
    \hline
    $\mu$ & molecular weight & 0.023~kg/mol (23~g/mol)\\
    $L$ & vaporization enthalpy  & 96.96~kJ/mol (9.696$\times$10$^{11}$~erg/mol) \citep{Fink-Leibowitz-1995:thermodynamic}\\ 
    $C_p$ & heat capacity  & 903.3~J/mol/K (9.033$\times$10$^{9}$~erg/mol/K) \\
    $A_{\mathrm{chem}}$ & parameter for chemical equilibrium pressure & 10$^{9.6}$~Pa (10$^{10.6}$~Ba) \citep{Castan-Menou-2011:atmospheres}\\
    $B_{\mathrm{chem}}$ & parameter for chemical equilibrium pressure & 38000K \citep{Castan-Menou-2011:atmospheres}\\
    $A_{\mathrm{sat}}$ & parameter for saturated vapor pressure & $10^{9.54}$~Pa (10$^{10.54}$~Ba) \citep{Bowles-Rosenblum-1965:vapor} \\
    $B_{\mathrm{sat}}$ & parameter for saturated vapor pressure & 12070.4K \citep{Bowles-Rosenblum-1965:vapor}\\
    $\rho_{\mathrm{droplet}}$ & density of sodium droplet & 968~kg/m$^3$ (0.968~g/cm$^3$) \\
    $A_{\mathrm{droplet}}$ & reflectivity of sodium droplet & 0.1 \citep{Barnett-Gentry-Jackson-et-al-1986:emissivity}\\
    $c_{\mathrm{Na}}$ & sodium concentration in mantle and crust & 0.29\% \citep{Schaefer-Lodders-Fegley-2012:vaporization}\\
    \hline
    \multicolumn{3}{l}{SiO-dominant atmosphere}\\
    \hline
    $\mu$ & molecular weight & 0.044~kg/mol (44~g/mol)\\
    $L$ & vaporization enthalpy & 411.5~kJ/mol (4.115$\times$10$^{12}$~erg/mol) \citep{Fink-Leibowitz-1995:thermodynamic}\\ 
    $C_p$ & heat capacity & 661~J/mol/K (6.61$\times$10$^9$~erg/mol/K) \\
    $A_{\mathrm{chem}}$ & parameter for chemical equilibrium pressure & 10$^{14.086}$~Pa (10$^{15.086}$~Ba) \\
    $B_{\mathrm{chem}}$ & parameter for chemical equilibrium pressure & 70300K\\
    $A_{\mathrm{sat}}$ & parameter for saturated vapor pressure & $10^{13.1}$~Pa ($10^{14.1}$~Ba)  \citep{Bowles-Rosenblum-1965:vapor} \\
    $B_{\mathrm{sat}}$ & parameter for saturated vapor pressure & 49520K \citep{Bowles-Rosenblum-1965:vapor}\\
    \hline
  \end{tabular}
  \caption{Parameter definitions.}
  \label{tab:parameters}
\end{table}

Finally, we want to highlight the assumptions we have made in our idealized model.

In the horizontal transport model, we completely ignore the rotation of the planet. Following \citet{Ingersoll-Summers-Schlipf-1985:supersonic}, we assume that the transport flow is in hydrostatic balance\footnote{The hydrostatic assumption is not used in the vertical escaping model, as the whole purpose is to calculate the vertical acceleration of the flow.}. This indicates that no radial escape flow will occur and that the mass flux of the transport flow will only be changed by condensation and surface exchange. Turbulence mixing in the boundary layer is ignored, as its effects are likely to be negligible \citep{Ingersoll-1989:io, Kite-Fegley-Schaefer-et-al-2016:atmosphere}. 

On the day-side, we assume that sodium dusts would revaporized by the stellar radiation immediately; while on the night-side, we assume that dusts, once formed, will stop interacting with the rest of the gas and enter a Kepler orbit determined by their initial energy and angular momentum, leading to a drop in pressure and mass flux. Actually, the different assumptions would only affect the flow property near the surface, where most condensation occurs. While the final escape fluxes would only be subtly affected, if at all. This is because the escape fluxes are mostly determined by the continuity conditions (Eq.~\ref{eq:transonic-eq-dry}-\ref{eq:transonic-cond-moist}) around the transonic point, where most condensation has already occurred.

We set $P$ and $T$ at the bottom boundary in the day-side vertical escape model using the surface temperature and corresponding equilibrium pressure. However, as we found in the horizontal transport model, the surface pressure is actually below the equilibrium pressure set by the magma ocean and vaporization relies on this subsaturation. This assumption could lead to overestimation of escape flux. Meanwhile, we also ignore the extra pressure induced by the stellar radiation and stellar wind on the day-side, and that could suppress the supersonic escape flow to a breeze as shown by \citet{Murray-Clay-Chiang-Murray-2009:atmospheric}.

The bottom boundary condition for the night-side escape model is set to conserve the mass and energy flux from the horizontal transport model. The assumption behind is that all energy would take form of either internal energy or mean flow kinetic energy (very small compared to internal energy in most cases), and no energy has be wasted on turbulence. This may be relevant here because the initial flow speed is quite slow, and possibly that will give the flow plenty of time to damp turbulence into internal energy. The cross section of the escape flow involves a large uncertainty too. We here arbitrarily stop the horizontal flow and let them turn upward when the flow speed drops $50\%$ from the peak value. A different choice could significantly change the results for cases with strong gravity, because, in those cases, lots of mass flux get lost due to the supersaturation in the small cross section. We also ignore the radiative cooling over time. This effect should be weak as a sodium vapor around 1000K is not hot enough to emit through its major spectrum lines.

Particle entrainment is not considered in the calculation, but we don't expect that will significantly change the results given that the initial velocity is typically small and thus the entrained particles at the surface are likely to fall back to the surface under gravity. What is also not considered is the chemical reaction, FUV driven ionization and heating. They could provide extra heat to the escape flow and enhance the escape flux as a result.

\bibliographystyle{aasjournal}



\end{document}